\documentclass[lettersize,journal]{IEEEtran}
\usepackage{amsmath,amsfonts}
\usepackage{algorithmic}
\usepackage{algorithm}
\usepackage{array}
\usepackage[caption=false,font=normalsize,labelfont=sf,textfont=sf]{subfig}
\usepackage{textcomp}
\usepackage{stfloats}
\usepackage{url}
\usepackage{verbatim}
\usepackage{graphicx}
\usepackage{xcolor}
\usepackage{cite}
\usepackage{svg}
\usepackage{booktabs}
\usepackage{multirow}
\usepackage[normalem]{ulem}
\usepackage{xcolor}

\newif\ifshowblue
\newif\ifshowred

\showbluefalse  
\showredfalse  

\ifshowblue
    \newcommand{\del}[1]{{\color{blue}\sout{#1}}}
\else
    \newcommand{\del}[1]{}
\fi

\ifshowred
    \newcommand{\add}[1]{{\color{red} #1}}
    \newcommand{\addtable}{\color{red}}
\else
    \newcommand{\add}[1]{#1}  
    \newcommand{\addtable}{}
\fi

\begin{document}

\title{Extended Field-of-View Passive mmWave Imaging Using Implicit Neural Representations}

\author{Sanjaya Herath,
        Matthew R. Ziemann,
        and~Christopher~A. Metzler%
\thanks{Sanjaya Herath is with the Department of Electrical and Computer Engineering, University of Maryland, College Park, MD 20742 USA (e-mail: sanjayah@umd.edu).}
\thanks{Matthew R. Ziemann is with DEVCOM Army Research Laboratory, Adelphi, MD 20783 USA (e-mail: matthew.r.ziemann.civ@army.mil).}
\thanks{Christopher A. Metzler is with the Department of Computer Science, University of Maryland, College Park, MD 20742 USA (e-mail: metzler@umd.edu).}

}

\maketitle

\begin{abstract}

We introduce a self-supervised learning-based framework for expanding the field-of-view (FoV) of passive \add{real-aperture} millimeter wave (mmWave) imaging systems. 
Our framework exploits aliasing artifacts/grating lobes---which are an inherent part of mmWave image formation---to expand FoV.  
Specifically, we recognize that removing aliasing artifacts can be cast as an underdetermined deconvolution problem and that in solving this problem, one can expand the system's FoV. 
We solve this deconvolution problem and expand the system's FoV by enforcing temporal regularity with time-varying Implicit Neural Representations (INRs). 
We demonstrate the effectiveness of our approach through extensive simulations and real-world experiments with Phase Sensitive Innovation's (PSI) state-of-the-art electro-optical modulation based passive mmWave imaging system.

\end{abstract}


\section{Introduction}
\label{sec:introduction}

\IEEEPARstart{M}{illimeter‐wave} (mmWave) systems, which operate at wavelengths on the order of millimeters (30–300 GHz), penetrate obscurants such as fog, smoke, and dust far more effectively than visible light. Moreover, mmWave signals can traverse certain materials---e.g., clothing and packaging---that are opaque at optical frequencies. As a result, mmWave imaging has emerged as a powerful tool for navigating and sensing under challenging low‐visibility conditions.

\add{Passive real-aperture millimeter-wave imaging systems operate by discretely sampling a continuous aperture field via antennas positioned at a sparse, fixed set of locations.}
\del{Millimeter-wave imaging systems operate by discretely sampling the continuous aperture field via antennas positioned at a sparse, fixed set of locations.} The resulting measurements are then processed via beamforming to estimate incoming power at multiple view directions. This sampling process makes mmWave systems susceptible to aliasing artifacts when the effective sampling interval, which is determined by the antenna spacing, exceeds the Nyquist limit for sampling \cite{aliasing_ref2}. When this condition is not met, high spatial frequency components of the scene fold into lower frequencies and produce aliasing artifacts or grating lobes. These artifacts get superimposed in the true image as ghost artifacts or duplicate features, potentially leading to misinterpretation \cite{mmwave_aliasing_ref1}.

In this work, rather than suppressing these artifacts, 
we exploit them to recover information beyond the native FoV. \add{Recognizing that these artifacts encode information from outside the native FoV, we formulate wide-FoV reconstruction as an underdetermined deconvolution problem using an aliasing-aware image formation model.} To solve this underdetermined deconvolution problem, we establish a regularized maximum likelihood estimation problem and employ a time‐varying INR to impose temporal consistency across successive narrow‐FoV measurements. By training an INR to map spatial coordinates and time indices to scene intensity, we reconstruct the wide FoV scene using the narrow FoV measurements. This effectively expands the system's FoV without additional hardware sampling. We validate the proposed approach through extensive simulations and real‐world experiments using Phase Sensitive Innovations’ (PSI) state‐of‐the‐art electro‐optical modulation–based mmWave imaging system. 

\section{Related Work}
\subsection{mmWave Imaging}
Millimeter-wave imaging has gained significant attention due to its ability to penetrate various materials while providing high-resolution images, making it particularly useful for security and surveillance applications. Most deployed systems, including airport body scanners, are active imagers that transmit mmWave signals and reconstruct scenes from reflected returns \cite{wang2019review}. Early studies demonstrated their effectiveness in detecting concealed objects and their advantages in low-visibility conditions. Advancements in hardware have further enhanced imaging capabilities, with MIMO-SAR testbeds enabling three-dimensional reconstruction \cite{yanik2020development} and super-resolution systems leveraging commodity 60-GHz WiFi radios to achieve high-quality imaging through thin materials \cite{zhang2020mmeye }. In contrast, passive mmWave imaging relies solely on thermally generated emissions from the environment \cite{10225255, yujiri2003passive}. This work focuses exclusively on passive mmWave imaging and addresses the challenges posed by spatial aliasing.

\subsection{Grating Lobe and Aliasing Suppression}
In many practical mmWave systems, while sub half-wavelength inter-antenna spacing is theoretically feasible, cost, space, and coupling constraints often render it impractical, resulting in spatial aliasing artifacts \cite{antenna_design_challenges}. To mitigate these artifacts, recent efforts employ nonuniform or jittered sparse‐periodic arrays guided by k-space analysis \cite{kazemi2021kspace}, along with wavenumber domain layout optimization that maximizes the peak-sidelobe ratios \cite{huan2024nonuniform} for alias-free imaging. Sparse array synthesis via convex optimization \cite{wang2023compressive} and geometrically inspired configurations such as spiral layouts \cite{ding2024geometric} reduce sidelobes and improve image fidelity. Accurate near-field reconstructions utilize a fast zero-migration algorithm to efficiently suppress grating lobes without increasing array redundancy \cite{zhu2025fast}. Moreover, compressive-sensing and sparse Bayesian learning frameworks enhance robustness under sparse sampling by embedding accurate near-field operators for alias-free reconstruction \cite{bi2022compressive, tu2025dualband}. While these methods primarily mitigate aliasing through array geometry design or sampling strategy optimization, our approach focuses on reconstruction-level mitigation by exploiting temporal redundancy in the data, enabling aliasing mitigation without requiring changes to the underlying hardware. 

\subsection{Exploiting Aliasing}
Many prior works exploit aliasing to recover additional scene information. 
A common approach is to introduce controlled spatial carriers that shift high-frequency content into the observable spectrum. For example, structured illumination microscopy (SIM) uses structured illumination to shift high-frequency information into the passband, enabling resolution enhancement from multiple measurements~\cite{sim}. Similarly, Fourier ptychography~\cite{fourier_ptychography} captures multiple shifted frequency measurements under varying illumination and combines them via phase retrieval, while holography~\cite{holography} employs off-axis reference beams to achieve a comparable frequency-shifting effect.

Meanwhile, several works exploit naturally occurring aliasing without controlled modulation. 
Prior studies have demonstrated super-resolution from aliased imagery by leveraging multiple undersampled observations with unknown subpixel shifts, jointly estimating relative displacements and reconstructing a high-resolution image using spatial and frequency-domain constraints or subspace methods~\cite{aliased_sr1, aliased_sr2, aliased_sr3, aliased_sr4}. These approaches show that aliasing can serve as a source of recoverable high-frequency information rather than merely a degradation.

In our setting, aliasing arises from the undersampled aperture, which folds high-frequency information into the measured band. We extract this aliased information by exploiting temporal consistency, without assuming rigid motion, by enforcing this structure with INRs.

\subsection{Multiplexed Imaging}
Traditional wide FoV imaging requires expensive optics or high-resolution sensors, which increases cost and complexity. Multiplexed imaging offers a promising alternative by encoding and superimposing measurements from multiple spatial regions onto a single sensor, and then computationally separating the overlapping signals.
Prior works explore a range of encoding mechanisms to multiplex spatially distinct scene regions. Spatial encoding strategies, including shift-based encoding, subfield encoding, and aperture division designs, map different FoV regions into a shared measurement while preserving their separability~\cite{uttam2009optically, shah2017shift, shepard2015design}. Beyond conventional imaging, multiplexing has been applied to structured image capture, off-axis digital holography, and event-based vision~\cite{gragston2019multiplexed, kumar2022single, ilani2023three}. 
While these methods demonstrate the versatility of multiplexing, they often rely on explicitly designed encoding strategies and often require multiple measurements or specialized optical configurations. In contrast, our approach does not rely on controlled multiplexing or custom hardware. Instead, we exploit aliasing-induced superposition inherent to the imaging system and leverage temporal structure to computationally reconstruct a wide FoV scene.

\subsection{Implicit Neural Representations}
INRs parameterize signals using continuous functions rather than discrete samples. Instead of relying on pixel grids or voxel-based models, INRs use neural networks to map spatial coordinates to signal values, effectively approximating the signal as a function. Because they are resolution-independent, INRs enable infinite resolution sampling and more efficient memory usage, scaling with signal complexity rather than spatial resolution. This makes them a powerful tool for representing a wide variety of complex signals~\cite{sitzmann2024awesome}. The pioneering works of~\cite{park2019deepsdf} and~\cite{mescheder2019occupancy} first showed that INRs outperform traditional methods in parameterizing geometry. Since then, INRs and related methods have achieved state-of-the-art results across various applications~\cite{sitzmann2020implicit,mildenhall2021nerf,saito2019pifu, feng2023neuws,Radish, sh-sas, mohan_dinr}, including multiplexed infrared imaging~\cite{saragadam2024foveated}, \add{video representations~\cite{nerv,videoinr}, and dynamic scene representations~\cite{dnerf,nerfies}}.
In this work, we employ time-varying INRs to model the wide FoV scene and enforce temporal consistency across frames.

\section{Background and System Model} 
\label{sec:system_model}

\subsection{Passive mmWave Imaging with Electro-Optical Modulation}
\label{subsec:imaging_w_eom}

Beamforming is a fundamental part of mmWave image formation: Beamforming coherently combines the signals received across multiple antenna elements to estimate the incoming power at various view directions. 
The intensity of each pixel in a mmWave image corresponds to the incoming power in that particular view direction. 

Fully digital beamforming—which requires sampling and processing the output of every array element—incurs substantial computational and hardware costs. This challenge becomes increasingly demanding in modern large‐scale arrays, which can contain over 1000 elements. Even the most efficient digital beamforming algorithms, such as FFT-based beamforming, require $O(N\log N)$ operations, where $N$ is the number of antenna elements \cite{vasquez2025synergistic}---this can represent a massive computational burden for large arrays. Therefore, forming images with large‐scale mmWave arrays using fully digital beamforming is often impractical, especially in real‐time applications.

\begin{figure}[t]
    \centering
        \includegraphics[width=\columnwidth]{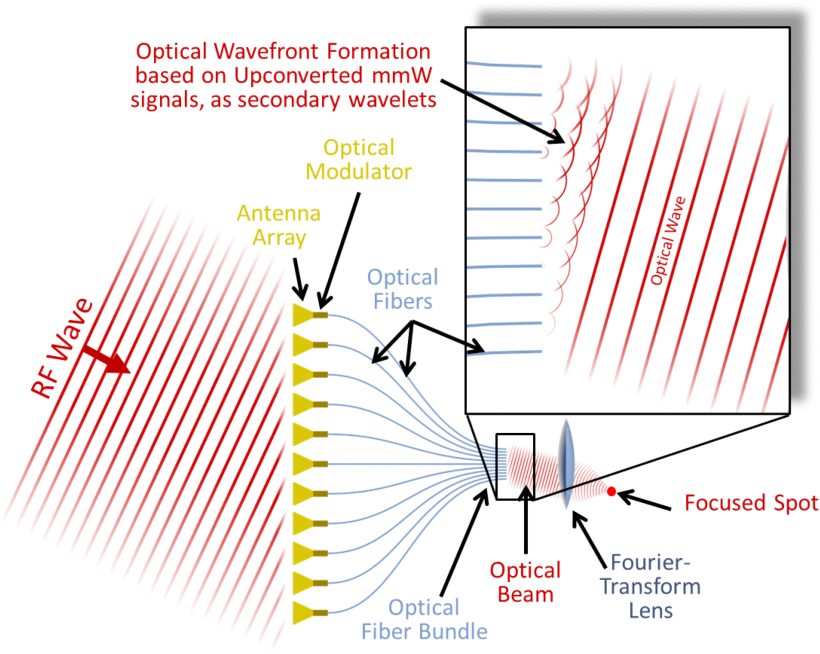}
        \caption{{\bf Upconversion process}. \add{The received RF signals are upconverted to the optical domain using electro-optical modulators (EOMs). An optical lens performs a two-dimensional spatial Fourier transform of the resulting optical field, mapping spatial frequencies to positions in the sensor array plane.} \del{The upconversion process uses electro‐optical modulators  (EOMs) to modulate the incoming RF signal and a lens to perform beamforming.} Adapted with permission from \cite{10225255}.}
        \label{fig:upconversion}
\end{figure}

\begin{figure}[t]
    \centering
        \includegraphics[width=.95\columnwidth]{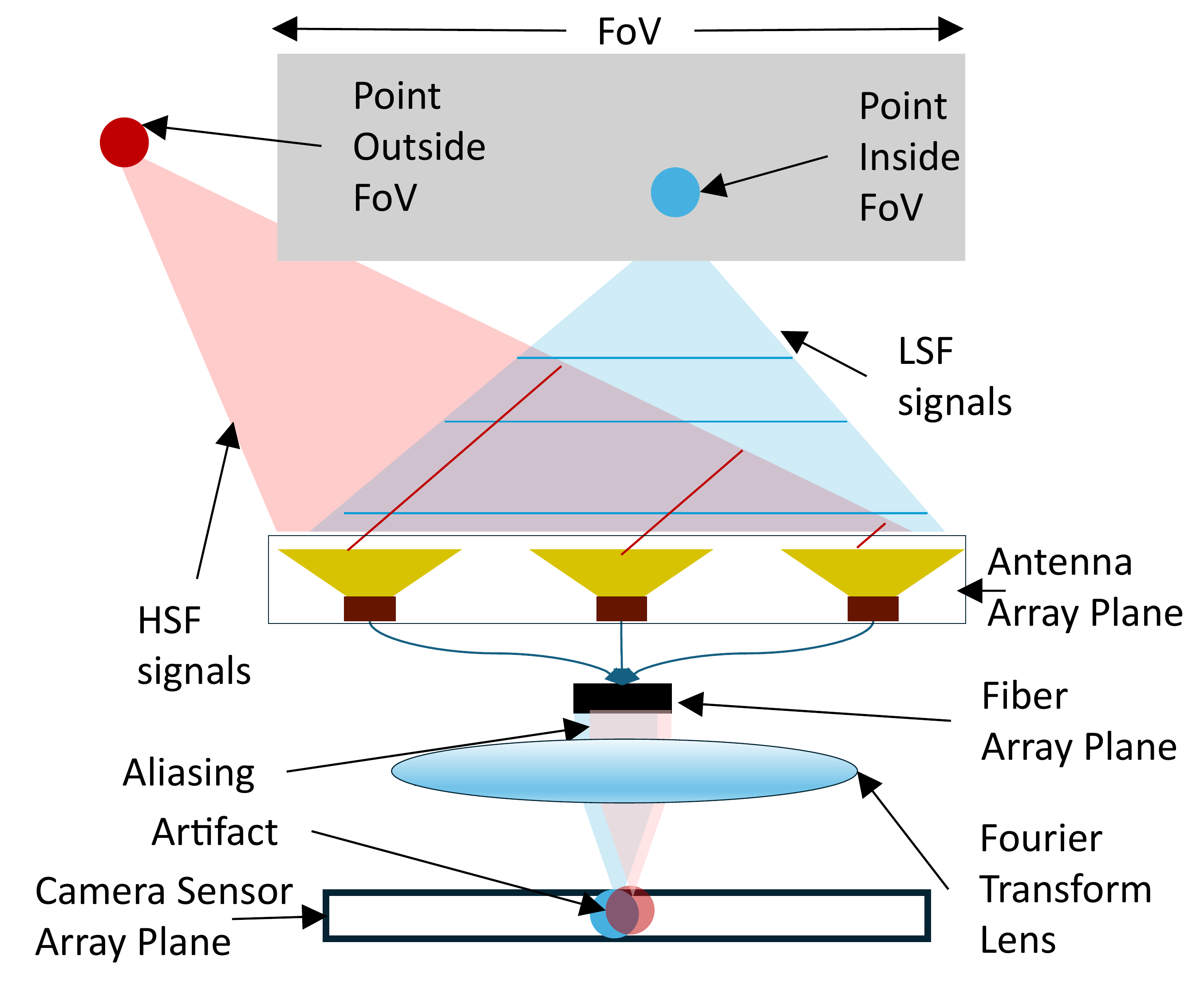}
        \caption{{\bf Aliasing process}. Points within the FoV correspond to low spatial frequency (LSF) signals on the antenna array and therefore do not produce aliasing. In contrast, points outside the FoV correspond to high spatial frequency (HSF) signals on the antenna array, leading to the formation of aliasing artifacts by super-imposing the HSF signals on the captured images.}
        \label{fig:twin_image}
\end{figure}

\begin{figure}[t]
    \centering
        \includegraphics[width=\columnwidth]{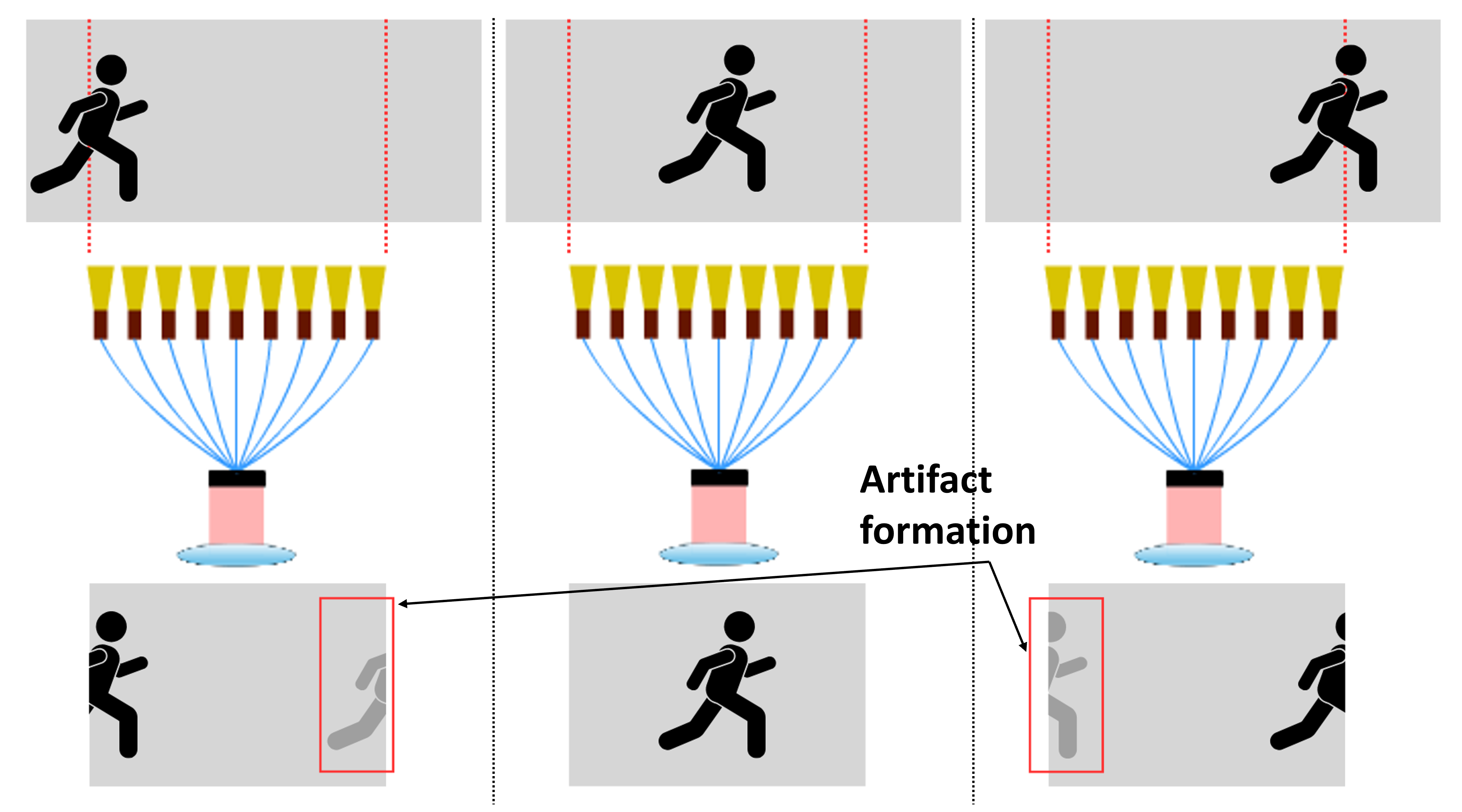}
        \caption{{\bf Artifact formation}.  When the object is within the FoV, upconversion introduces no artifacts; when it is outside, aliasing artifacts form. Note: Sensor image visualization has been reflected for clarity. }
        \label{fig:artifact_formation}

\end{figure}

Phase Sensitive Innovations (PSI) developed an alternative to fully digital beamforming by upconverting mmWave signals into the optical domain \cite{10225255}. High‐speed electro‐optical modulators (EOMs) sample the incoming RF wavefronts—preserving both amplitude and phase—and map them onto an optical carrier. \protect\del{In the optical domain, a simple lens can perform a spatial Fourier transform, enabling instantaneous beamforming over a continuum of viewing angles without per‐element digitization.}
\add{The modulated optical signals
are transmitted through optical fibers and relaunched into
free space from a fiber array. An optical lens then performs a two-dimensional spatial Fourier transform of the resulting optical field, mapping spatial frequency components to positions in the camera sensor plane and enabling instantaneous beamforming over a continuum of viewing angles without per-element digitization
\cite{goodman2005introduction}. The camera records the
resulting intensity distribution, producing a mmWave image
in which sensor positions correspond to viewing directions.}
By offloading beamforming to this hybrid electro‐optical process, the system maintains high angular resolution and real‐time operation while reducing computational complexity from $O(N\log N)$ to $O(1)$. Fig.~\ref{fig:upconversion} illustrates the complete upconversion and optical beamforming workflow.

\subsection{Spatial Aliasing and Grating Lobes}

\del{
Similar to digital mmWave imaging systems, electro-optical mmWave imaging system experiences aliasing effects when one undersamples the optical aperture.  
This undersampling gives rise to optical grating lobes in the image plane. Additional discussion of the origin and impact of optical grating lobes in PSI’s system is provided in cite{psi aliasing}. This behavior is equivalent to the spatial aliasing observed in discretely sampled RF arrays. 
In both cases, insufficient spatial sampling redistributes energy into unintended angular directions, producing aliasing artifacts. Consequently, objects located outside the antenna array’s field of view (FoV) can still induce responses on the sampled aperture, and these {out-of-FoV} contributions superimpose onto the desired image, leading to potential misinterpretation. 
}

\del{
This aliasing behavior can be further understood by examining how the spatial frequency content of the scene interacts with the sampling limits of the aperture. When an object lies within the designed FoV, its wavefront generates spatial frequencies at the aperture that remain below the Nyquist limit, so no aliasing occurs. However, objects beyond the FoV introduce higher spatial frequencies that exceed this Nyquist threshold. Sampling these high-frequency components folds them back into the passband, producing aliased replicas that overlap the true image, thereby creating the aliasing artifacts illustrated in Fig.~\ref{fig:twin_image}. An example of the aliasing artifacts and their effect on image interpretation is illustrated in Fig.~\ref{fig:artifact_formation}.
}

\add{
Spatial aliasing in the considered imaging system originates from discrete spatial sampling of the incoming RF field at the antenna array plane. When the antenna spacing is sufficiently large, signals arriving from different directions can produce identical sampled phase progressions, resulting
in grating-lobe ambiguities. 

This ambiguity can also be understood in the spatial-frequency domain. The direction of an incoming wave determines its spatial frequency across the antenna aperture. When these spatial frequencies exceed the sampling Nyquist limit, they fold into lower spatial frequencies and become indistinguishable from signals arriving from other directions. This behavior is illustrated in
Fig.~\ref{fig:twin_image}.

Because the electro-optical conversion preserves the relative phases of the sampled RF signals, the spatial aliasing ambiguity is carried into the optical domain. The optical lens subsequently performs Fourier-transform beamforming,
and the resulting grating-lobe responses can map out-of-FoV contributions into the measured FoV. Consequently, radiation originating outside the native FoV
can overlap with in-FoV contributions, producing ghost artifacts in the captured image. An example of this artifact formation is illustrated in
Fig.~\ref{fig:artifact_formation}.
}

\subsection{Imaging Forward Model}

\del{
The imaging process described above can be represented using an aliasing-aware forward model that relates the underlying wide-FoV scene to the measured narrow-FoV image. Specifically, we recognize that mmWave image formation can be modeled according to
}
\add{As discussed in the previous subsection, spatial aliasing causes out-of-FoV contributions to overlap with the measured image. To capture this effect, we model the relationship between the underlying wide-FoV scene and the measured narrow-FoV image using the following aliasing-aware forward model:}
\begin{equation}\label{eqn:Forward}
    \mathbf{y}_t = \mathbf{M}\!\left(h * \mathbf{x}_t + \mathbf{n}_t\right), 
    \quad t = 1, 2, \dots, T,
\end{equation}

\noindent where $\mathbf{y}_t \in \mathbb{R}^{H_y \times W_y}$ denotes the measured narrow–FoV image, $h$ denotes the system point spread function (PSF), $\mathbf{x}_t \in \mathbb{R}^{H_x \times W_x}$ is the underlying wide–FoV scene with $H_y \leq H_x$ and $W_y \leq W_x$, and $\mathbf{n}_t$ denotes additive noise. \add{$\mathbf{M}: \mathbb{R}^{H_x \times W_x} \rightarrow \mathbb{R}^{H_y \times W_y}$ represents a deterministic spatial cropping operation that extracts the central $H_y \times W_y$ region of the convolved wide-FoV image, corresponding to the sensor's native FoV.}

\add{
The PSF $h$ characterizes the system's response to a point source, including the grating lobes introduced by spatial undersampling. We derive its analytical form and introduce a practical approximation in the following subsection.
}

\subsection{PSF Characterization}

\begin{figure}[t]
    \centering
    \addtable
    \includegraphics[width=.5\textwidth]
        {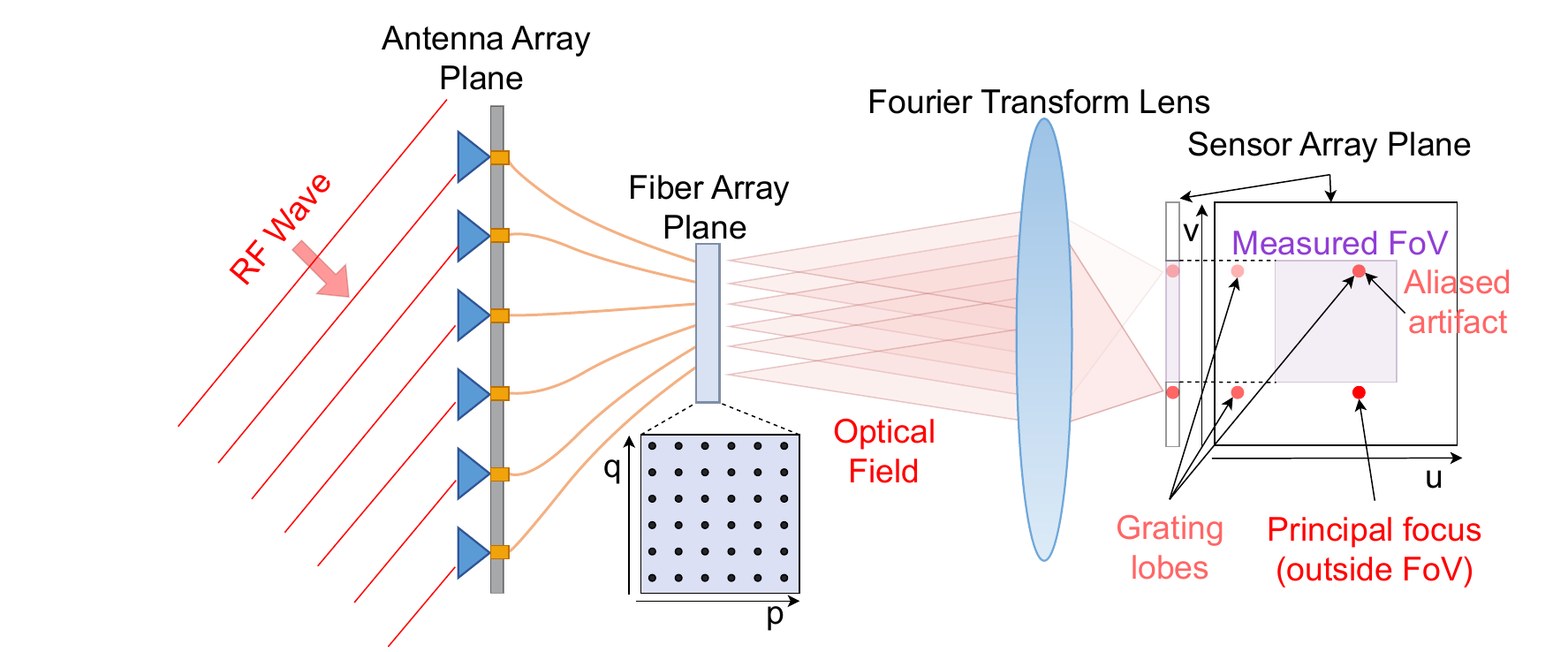}
    \caption{\textbf{mmWave image formation and spatial aliasing.}
    The incoming mmWave field from a far-field point source is spatially sampled by the antenna array and upconverted to the optical domain using EOMs.
    The optical signals are transmitted through fibers and relaunched into free space at the fiber array plane. A Fourier-transform lens maps the resulting optical field onto the camera focal plane, where the intensity distribution
    is recorded. Due to the periodic spatial sampling of the aperture,
    the intensity distribution contains a principal focus and grating-lobe replicas. For a source outside the nominal FoV, the principal focus lies outside the observed region, while grating-lobe replicas can fall within the sensor's FoV, producing aliasing artifacts in the measured image.}
    \label{fig:mmwave_image_formation}
\end{figure}

\del{To understand the origin of aliasing artifacts, we first analyze the system’s PSF, which characterizes how a point source is mapped through the imaging process. 
The EOMs will capture the point source and upconvert it. At the lens, it will be represented as a grid of point sources. To model this, first we consider an $n \times n$ grid of point sources, where each point source is represented by a Dirac delta function. This can be expressed as:}

\add{
To characterize the aliasing behavior described above, we analyze the system's PSF, which describes its response to a point source. As illustrated in Fig.~\ref{fig:mmwave_image_formation}, the antenna array spatially samples the incident RF wavefront, and the received signals are subsequently upconverted to the optical domain using EOMs while
preserving their relative phase information.
The optical signals are then relaunched from the fiber array, forming a discrete optical aperture whose field undergoes a spatial Fourier transform
through the lens.
To model this process, we represent the fiber array plane as a two-dimensional grid of discrete optical sources on an $n \times n$ computational grid, where each source is represented by a discrete impulse. This can be expressed as

\begin{equation}
S[p,q] =
\sum_{i=0}^{Q-1}\sum_{j=0}^{Q-1}
\delta[p-iR]\delta[q-jR],
\label{eq:samplingmask_def}
\end{equation}
\noindent where $p,q=0,1,\ldots,n-1$ are the spatial
indices of the computational grid, $Q=n/R$, $R$ denotes the spacing between adjacent fiber outputs in the computational grid units, and $\delta[\cdot]$ is the Kronecker delta function.
}

Next, to extend this grid to incorporate an extended FoV, we center the grid at the origin to consider a $2n \times 2n$ window. This extended grid can be expressed as

\add{
\begin{equation}
P[k,l] =
\begin{cases}
S[k-\frac{n}{2},l-\frac{n}{2}],
& \frac{n}{2}\leq k,l<\frac{3n}{2},\\
0, & \text{otherwise}.
\end{cases}
\label{eq:extended_grid_def}
\end{equation}
}
\noindent where \( k,l=0,1,\ldots,2n-1 \). All non-zero entries of $P[k,l]$ lie along a grid of points inside the central block of the padded array, spaced by $R$ in both directions, starting at index $\frac{n}{2}$.

\del{
Computing the discrete Fourier transform (DFT) of the extended grid then yields the PSF of the system:}
\add{As discussed in section~\ref{subsec:imaging_w_eom}, the optical lens performs a spatial Fourier transform of the field relaunched from the fiber array. Therefore, we compute the discrete Fourier transform (DFT) of the aperture
distribution $P[k,l]$ to obtain the corresponding coherent PSF at the camera sensor array plane:}
\begin{equation}
    \text{PSF}(u,v) = \frac{1}{(2n)^2}\sum_{k=0}^{2n-1} \sum_{l=0}^{2n-1} \text{P}[k,l]\,
    e^{j2\pi\left(\frac{k}{2n}u + \frac{l}{2n}v\right)},
\label{eq:psf_initial}
\end{equation}

\noindent where \( u,v = 0,1,\ldots,2n-1 \) denote the pixel indices in the \add{camera} sensor \add{array} plane.

From Eq.~\eqref{eq:extended_grid_def}, the non-zero entries of $P[k,l]$ occur at $k=\frac{n}{2}+iR$ and $l=\frac{n}{2}+jR$,
where $i,j=0,1,\ldots,Q-1$.
Substituting these indices into Eq.~\eqref{eq:psf_initial} yields
\begin{equation}
    \text{PSF}(u,v) = \frac{1}{(2n)^2}
    \sum_{i=0}^{Q-1} \sum_{j=0}^{Q-1}
    e^{j2\pi\left(
    \frac{(n/2+iR)}{2n}u +
    \frac{(n/2+jR)}{2n}v
    \right)}.
\label{eq:psf_second}
\end{equation}

\noindent This expression can be factored into a product of two summations:
\begin{equation}
    \text{PSF}(u,v)
    = \frac{1}{4n^2}
    e^{j\pi n\frac{u+v}{2}}
    \left(\sum_{i=0}^{Q-1} e^{j\pi\frac{Ri\,u}{n}}\right)
    \left(\sum_{j=0}^{Q-1} e^{j\pi\frac{Rj\,v}{n}}\right).
\label{eq:psf_third}
\end{equation}

\noindent Let the first summation be denoted by \( S_u(u) \) and the second by \( S_v(v) \).
We now consider \( S_u(u) \). Defining
\[
\alpha = \frac{\pi R u}{n},
\]
we obtain
\begin{equation}
    S_u(u) = \sum_{i=0}^{Q-1} e^{j\alpha i}.
\end{equation}

\noindent This is a geometric series with \( Q \) terms and common ratio \( e^{j\alpha} \).
When \( \alpha \bmod 2\pi = 0 \), the summation simplifies to \( Q \).
For \( \alpha \bmod 2\pi \neq 0 \), the summation admits the closed-form expression
\begin{equation}
    S_u(u)
    = \frac{1 - e^{j\alpha Q}}{1 - e^{j\alpha}}
    = e^{j\alpha\frac{Q-1}{2}}
      \frac{\sin\!\left(\frac{Q\alpha}{2}\right)}
           {\sin\!\left(\frac{\alpha}{2}\right)}.
\end{equation}

\noindent An identical expression holds for \( S_v(v) \). Consequently, the \add{coherent} PSF can be written as
\begin{equation}
    \text{PSF}(u,v)
    = \frac{1}{4n^2}
      e^{j\pi n\frac{u+v}{2}}
      S_u(u)\, S_v(v).
\label{eq:psf_fourth}
\end{equation}

\noindent As our system records the intensity of the incoming field, the system's response is characterized by the magnitude-squared of the coherent PSF. The resulting \add{intensity} PSF is given by
\begin{equation}
\label{eq:psf_ideal}
    \text{PSF}(u,v)
    = \frac{1}{16n^4}
      \left|
      \frac{
      \sin\!\left(\frac{\pi u}{2}\right)
      \sin\!\left(\frac{\pi v}{2}\right)
      }{
      \sin\!\left(\frac{\pi R u}{2n}\right)
      \sin\!\left(\frac{\pi R v}{2n}\right)
      }
      \right|^2.
\end{equation}

While Eq.~\eqref{eq:psf_ideal} characterizes the idealized system response, it does not fully capture the behavior observed in practice. Real mmWave hardware exhibits non-uniform aperture illumination, which suppresses grating lobes relative to the main lobe, and systems employing EOMs additionally suffer from spatial blurring due to imperfect focusing and optical aberrations. To obtain a more realistic forward model, we incorporate both aperture tapering and Gaussian blur into the PSF.

To model aperture tapering, we apply a 2-D Kaiser window with parameter \( \beta = 4 \). Let \( W(u,v) \) denote the separable Kaiser window,
\begin{equation}
W(u,v)
= w_{\text{Kaiser}}(u;\beta)\, w_{\text{Kaiser}}(v;\beta),
\end{equation}
where \( w_{\text{Kaiser}}(\cdot;\beta) \) is the standard 1-D Kaiser window.

To model spatial blurring, we convolve the tapered PSF with a 2-D Gaussian kernel of standard deviation \add{\( \sigma_G = 1 \)}. Let \( h_{\text{G}}(u,v) \) denote this Gaussian kernel.

The resulting practical PSF is therefore
\begin{equation}
\label{eq:psf_real}
\text{PSF}_{\text{real}}(u,v)
=
h_{\text{G}} * \left[ W(u,v)\,\text{PSF}(u,v) \right],
\end{equation}
where \( * \) denotes 2-D convolution.

\add{The parameters $\beta$ and $\sigma_{\mathrm{G}}$ were selected empirically based on the reconstruction quality of real-world measurements (See Appendix~\ref{subsec:psf_ablation}). }

\begin{figure*}[t]
    \centering
        \includegraphics[width=\linewidth]{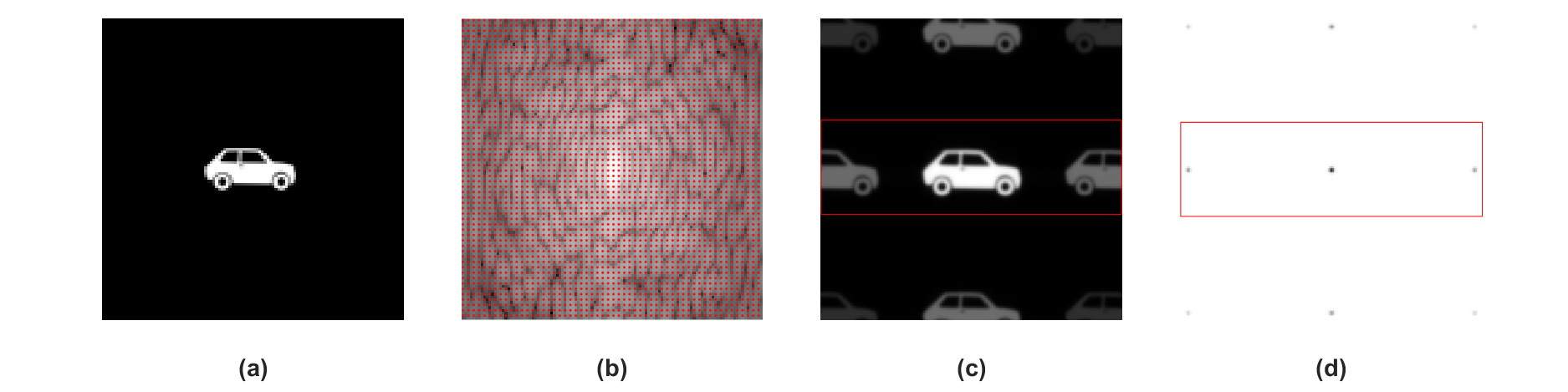}
        \caption{{\bf Measurement of a far field scene with the modeled analytical point spread function}. (a) Far field scene. (b) Field distribution at the aperture with red dots indicating the antenna locations. (c) Measured scene with the aliasing artifacts (the red bounding box indicates the FoV). (d) modeled PSF (displayed inverted for better visual contrast). The red bounding box indicates the region of interest for the PSF in this work. }
        \label{fig:psf_model}
\end{figure*}

 We set $n=112$ and $ R=2$ to obtain the PSF that aligns with measurements from the mmWave imaging system. Fig.~\ref{fig:psf_model} illustrates the PSF resulting from Eq.~\eqref{eq:psf_real} and how a far field scene is modeled by the PSF. In this paper, we restrict our analysis to the horizontal aliasing artifacts, which are widely noticeable in the mmWave imaging data. For that, we consider the region bounded by the red bounding box in Fig.~\ref{fig:psf_model}-(d).

\subsection{Noise Characterization}

We conducted a detailed analysis of the system noise to inform our modeling assumptions (see Appendix~\ref{sec:appendix a} for details). The analysis was performed using 200 RAW frames of a static scene to isolate instrument-induced variations. Statistical evaluation via quantile–quantile (QQ) plots confirmed that the noise is approximately Gaussian distributed. Power spectral density (PSD) analysis revealed a broadly flat spectrum, indicating weak spatial correlation and near-whiteness. Temporal correlation analysis further showed minimal inter-frame dependence. Collectively, these results justify modeling the system noise as white Gaussian noise in our formulation.

\section{INR-Based Reconstruction}
\label{sec:extended_fov}

Building upon the forward model and noise characteristics established in Section~\ref{sec:system_model}, we formulate image reconstruction as a maximum likelihood estimation problem, which we regularize using INRs. Our framework integrates two key components: a likelihood-based loss function derived from the imaging process and an INR-based parameterization of the wide FoV scene to enforce smoothness and spatio-temporal consistency.

\subsection{Log-Likelihood Loss} 

To reconstruct the wide FoV scene, we define a loss function for the INR such that it approximates intensity values of the wide FoV scene to maximize the likelihood of the observed narrow FoV measurements. To obtain the loss function, we proceed as follows.

We first reformulate the imaging process in Eq.~\eqref{eqn:Forward} as,

\begin{equation}
    \mathbf{y}_t = F(\mathbf{x}_t) + \mathbf{M}\mathbf{n}_t,
\end{equation}
where \( F(\mathbf{x}_t) = \mathbf{M}\!\left(h * \mathbf{x}_t\right) \) and we approximate the noise \( \mathbf{n}_t \) as \add{independent and identically distributed} Gaussian random variable with zero mean and variance \( \sigma^2 \), which is consistent with the noise characterization (see Appendix~\ref{sec:appendix a}). 
\del{Since $\mathbf{M}$ is a binary mask, the masked noise $\mathbf{M}\mathbf{n}_t$ remains zero-mean Gaussian with variance $\sigma^2$ on the observed pixels.}
\add{Since $\mathbf{M}$ is a deterministic spatial cropping operator that selects pixels without modifying their values, the cropped noise $\mathbf{M}\mathbf{n}_t$ remains zero-mean Gaussian with variance $\sigma^2$ on the observed pixels.}
 Treating each pixel \((u,v)\) as an independent Gaussian observation, the likelihood of observing $\mathbf{y}_t$ given a candidate wide‑FoV scene $\mathbf{x}_t$ is:

\begin{equation}
    p\!\left(\mathbf{y}_t \mid \mathbf{x}_t\right)
    =
    \prod_{u=1}^{W_y} \prod_{v=1}^{H_y}
    \frac{1}{\sqrt{2\pi\sigma^2}}
    \exp\!\left[
        -\frac{
            \left\|
                (\mathbf{y}_t)_{u,v}
                -
                (F(\mathbf{x}_t))_{u,v}
            \right\|^2
        }{2\sigma^2}
    \right].
\end{equation}

The negative log-likelihood of the observed narrow FoV measurements $\mathbf{y}_t$ given the candidate wide FoV scene $\mathbf{x}_t$ can be expressed as:

\begin{equation}
    -\log p\!\left(\mathbf{y}_t \mid \mathbf{x}_t\right)
    =
    \sum_{u=1}^{W_y} \sum_{v=1}^{H_y}
    \frac{
        \left\|
            (\mathbf{y}_t)_{u,v}
            -
            \bigl(F(\mathbf{x}_t)\bigr)_{u,v}
        \right\|^2
    }{2\sigma^2}
    + \mathrm{const}.
\end{equation}

\noindent Since the additive constant and the factor \(1/(2\sigma^2)\) do not affect the minimizer, we drop them and collapse the spatial sums into an \(\ell_2\) norm, yielding the final loss function:

\begin{equation}
    \label{eqn:loss}
    \mathcal{L} = \sum_{t=1}^{T} \|\mathbf{y}_t - F({\mathbf{x}_t})\|^2,
\end{equation}

\noindent which is the  \(\ell_2\) loss between the observed narrow FoV measurements \( \mathbf{y}_t \) and the narrow FoV scene \( F({\mathbf{x}_t}) \). Therefore, we use an INR to parametrize the wide FoV estimate $\hat{\mathbf{x}}$ and train it by minimizing $\mathcal{L}$, ensuring that its narrow FoV estimates $F(\hat{\mathbf{x}}_t)$ are consistent with the observed measurements $\mathbf{y}_t$.

\subsection{Regularized Reconstruction with INRs}

We model and reconstruct the wide FoV scene estimate \( \hat{x}_t \) using an INR~\cite{sitzmann2020implicit}. INRs are a class of neural networks that represent data as continuous functions parameterized by neural networks; importantly, these parameterizations can be a function of both space and time. INRs naturally capture and exploit temporal regularity in the signals they represent~\cite{ziemann2024time}.

\begin{figure}[t]
    \centering
    \addtable
    \includegraphics[width=\columnwidth]{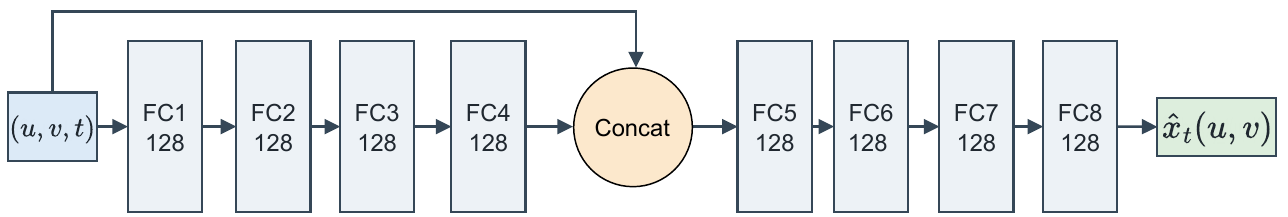}
    \caption{\textbf{INR Architecture}. An eight-layer MLP
    maps spatial and temporal coordinates $(u,v,t)$
    to scene intensity $\hat{x}_t(u,v)$.
    }
    \label{fig:INR_architecture}
\end{figure}

We design our INR as an eight-layer multilayer perceptron with 128 hidden units per layer. The input coordinate vector $(u,v,t)$ is processed through four hidden layers, after which the original input is concatenated with the intermediate feature representation via a single skip connection. This design allows the network to preserve low-level spatial--temporal information while enabling subsequent layers to model higher-order structure. ReLU activations are employed in all hidden layers. Dropout regularization is applied after selected activation layers, while the skip connection itself remains unregularized. 
\add{The architecture of the proposed INR is illustrated in Fig.~\ref{fig:INR_architecture}.}

 The INR takes in spatial coordinates $(u,v)$ and a time index $t$ and produces an estimate of the scene intensity at pixel $(u,v)$ at time $t$. That is, the scene estimate at time $t$ is given by
\begin{equation}
    \hat{\mathbf{x}}_t = f_{\theta}(\mathbf{z}_t),
\end{equation}
\noindent where \( f_{\theta} \) is the INR with trainable parameters \( \theta \) and $\mathbf{z}_t=\{(u, v, t) \}_{u=1, v=1}^{W_x, H_x}$ is the set of all pixel coordinates at time $t$. 

With this scene representation in hand, we can evaluate and minimize our prediction loss by minimizing the loss function defined in Eq.~\eqref{eqn:loss}. The loss function enforces consistency between the measured narrow FoV observations and the corresponding model predictions, serving as the primary driver for INR training. 
\del{We also introduce a total variation (TV) regularization term, $ \sum_t \mathrm{TV}(\hat{\mathbf{x}}_t)$, to encourage spatial smoothness while preserving edges.}
\add{Although the INR provides implicit regularization, we also introduce a total variation (TV) term, $\sum_t \mathrm{TV}(\hat{\mathbf{x}}_t)$, to suppress spatial artifacts while preserving edges.  This proves particularly important with real-world measurements, as demonstrated in Appendix~\ref{subsec:reg_ablation}.}
The final loss used for all experiments is therefore given by
\add{
\begin{equation}
    \label{eqn:loss_reg}
    \mathcal{L} = \sum_{t=1}^{T} \left(
    \|\mathbf{y}_t - F(\hat{\mathbf{x}}_t)\|_2^2
    + \lambda \,\mathrm{TV}(\hat{\mathbf{x}}_t)
    \right),
\end{equation}
where $\lambda$ controls the relative strength of the total variation regularization term.
}

\begin{figure*}[t]
    \centering
        \includegraphics[width=\textwidth]{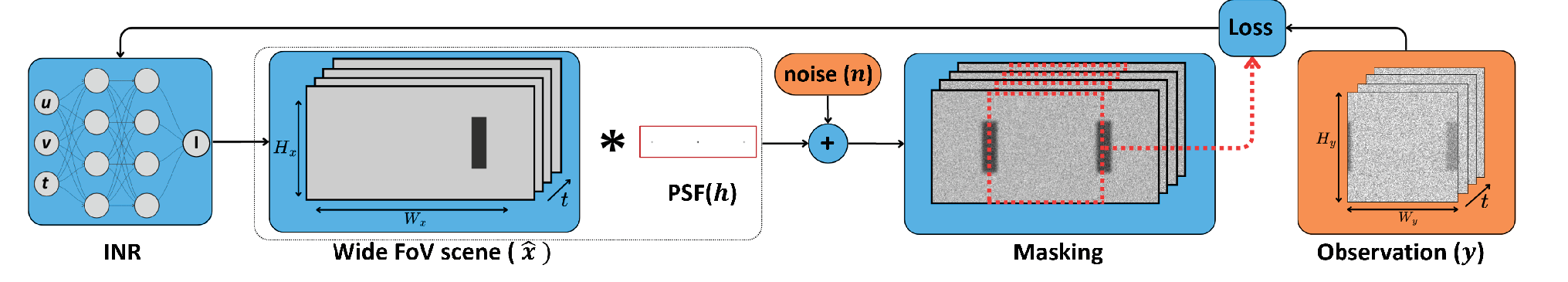}
        \caption{{\bf Proposed framework}. The Implicit Neural Representation (INR) receives spatial and temporal coordinates as input, and it outputs the estimated pixel intensity of the wide FoV scene (\( \hat{x} \)). The narrow FoV scene (\( y \)) is obtained by the imaging process described in Eq.~\eqref{eqn:Forward}. The parameters of the INR are optimized to minimize the loss function defined in Eq.~\eqref{eqn:loss_reg}. }
        \label{fig:framework}
\end{figure*}

The proposed INR framework is implemented in PyTorch~\cite{pytorch} and an overview of the method is illustrated in Fig.~\ref{fig:framework}. \add{Despite the spectral bias of conventional MLPs, the proposed framework exploits spatial aliasing and temporal consistency to recover extended-FoV information. Alternative INR architectures with enhanced high-frequency representation capabilities are evaluated in Appendix~\ref{subsec:inr_ablation}.}

\section{Experiments}

\subsection{Baseline Methods and Metrics}

To evaluate reconstruction performance, we compare the proposed method against five baseline approaches: 
Deep Video Prior (DVP)~\cite{deepVideoPrior}, 
a low-rank model \del{solved via alternating gradient descent (altGD)} \add{(AltGDmin)}~\cite{altGDmin}, \add{a low rank plus sparse model (L+S) \cite{lpluss},}
a U-Net model that performs frame-to-frame reconstruction (Unet-Frame)~\cite{unet}, 
and a U-Net model that uses a sequence of frames for sequence-to-frame reconstruction (Unet-Seq).

AltGDmin and L+S are model-based optimization methods that exploit
temporal correlations across the entire sequence. AltGDmin assumes a low-rank representation and alternately optimizes its factors, 
\add{while L+S decomposes the video into low-rank and sparse components.}
DVP uses a convolutional neural network (CNN) to exploit temporal coherence across an entire sequence. We also consider two U-Net variants:
Unet-Frame reconstructs individual frames, while Unet-Seq incorporates temporal context from a window of $L=10$ frames.

All methods reconstruct the scene without ground-truth wide-FoV images by enforcing measurement consistency through the same PSF-based forward model. The deep learning methods are trained in a self-supervised manner.
These baselines allow us to compare explicit structural priors, CNN-based representations, and different levels of temporal modeling against the proposed method. 
\add{Further details on the baseline methods are provided in Appendix~\ref{sec:supp_baselines}.}

\paragraph{Training details}
The neural network models are trained using the Adam optimizer, with a learning rate of $0.001$ for the proposed INR and both U-Net baselines, and $0.0001$ for DVP. 
\add{For AltGDmin, we employ a low-rank factorization with rank $r=20$ and alternating optimization, using Adam to update $\mathbf{U}$ with a learning rate of $0.01$. The L+S baseline is optimized using proximal gradient descent.}

For the INR and the U-Net baselines, a learning-rate scheduler reduces the learning rate by a factor of $0.1$ if the training loss does not improve for $10$ consecutive epochs. Training is performed for up to $6000$ epochs, with early stopping triggered after $50$ consecutive epochs without improvement. Similarly, AltGDmin and L+S are optimized for up to $6000$ iterations, with early stopping applied if the objective function does not improve for $50$ consecutive iterations.

\add{
The TV regularization term is included in the reconstruction objective for all methods except L+S, which employs its own nuclear-norm and sparsity regularization. Based on our ablation study, we set $\lambda=0$ for the simulated experiments, as explicit regularization did not provide consistent improvements in reconstruction quality. For real-data experiments, we set $\lambda=1$ to help suppress residual artifacts in the reconstructed scenes.
}
\add{
All experiments were conducted on a Linux machine equipped with an AMD EPYC 7H12 Rome CPU and an NVIDIA Tesla V100 GPU with 32 GB of memory. The models were implemented in  PyTorch.
}
\add{
\paragraph{Evaluation metrics}
For quantitative evaluation, we consider four widely used perceptual metrics: Peak Signal-to-Noise Ratio (PSNR), Structural Similarity Index Measure (SSIM), Learned Perceptual Image Patch Similarity (LPIPS) \cite{lpips}, and Fréchet Video Distance (FVD) \cite{fvd}. 
PSNR measures pixel-level reconstruction fidelity, while SSIM evaluates structural similarity. 
LPIPS measures perceptual similarity between images by comparing deep feature representations extracted from neural networks, providing a closer alignment with human visual judgment. FVD evaluates the distributional similarity between video sequences using pretrained spatiotemporal features, capturing both spatial quality and temporal dynamics. Higher PSNR and SSIM values indicate better reconstruction quality, whereas lower LPIPS and FVD values indicate better perceptual and video quality.
}

\begin{figure*}[h]
    \centering
        \includegraphics[width=\textwidth]{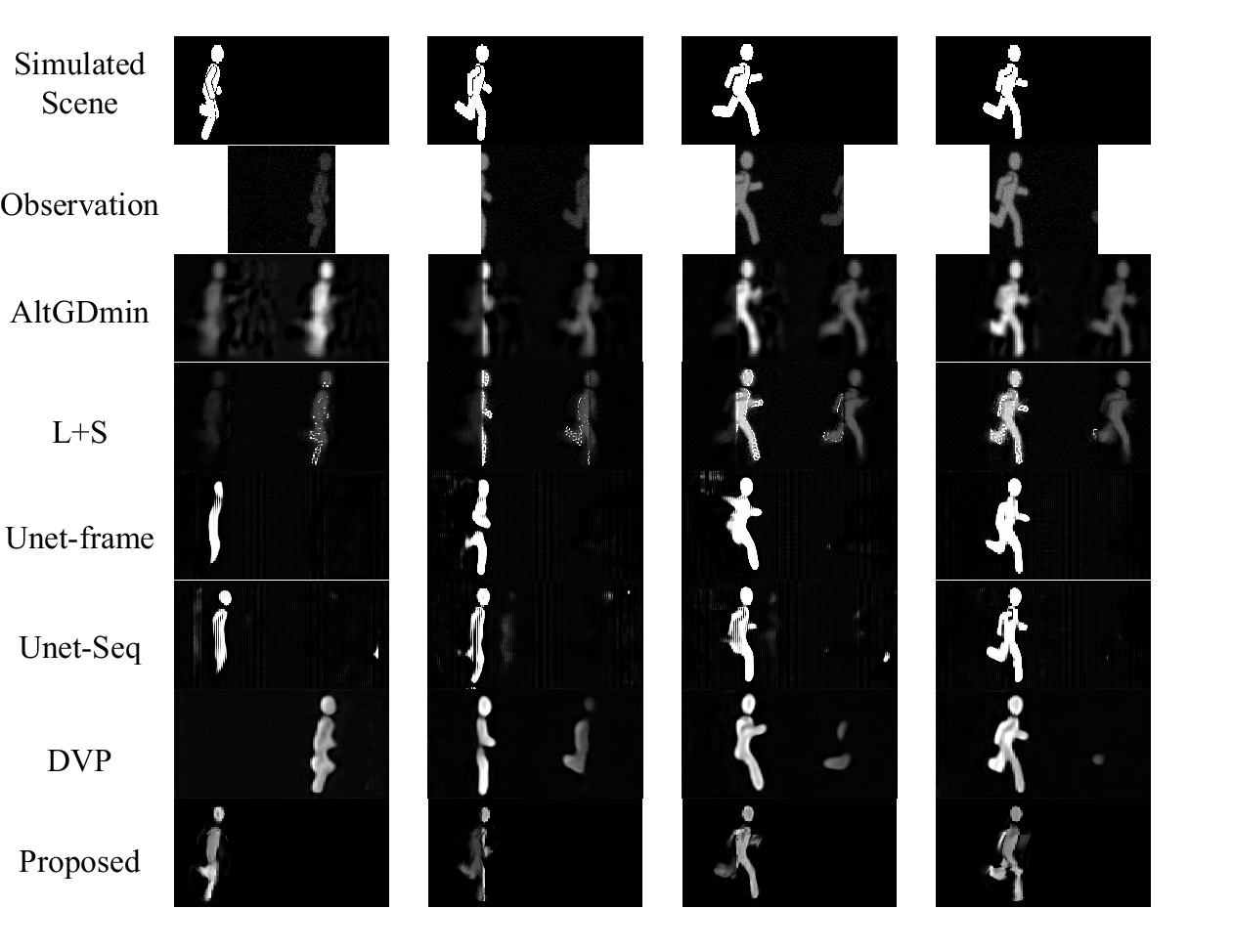}
        \caption{{\bf Extended FoV reconstruction of simulated non-rigid scene (running motion)}. Selected reconstructed frames for the simulated non-rigid data. The proposed method effectively reconstructs a wide FoV from narrow FoV measurements while removing the aliasing artifacts compared to the baseline methods. Please see the video supplement.}
        \label{fig:simulated_dynamic_results}
\end{figure*}

\begin{table}[t]
\centering
\addtable
\caption{Quantitative comparison of reconstruction performance
across three simulated sequences. Best results are in bold.}
\label{tab:sim_quant_results}

\resizebox{\columnwidth}{!}{%
\begin{tabular}{@{}clcccc@{}}
\toprule
Sequence & Method & PSNR (dB) $\uparrow$ &
SSIM $\uparrow$ & LPIPS $\downarrow$ &
FVD $\downarrow$ \\
\midrule

\multirow{6}{*}{\shortstack[c]{Rigid scene\\
(Fig.~\ref{fig:simulated_results})}}
& AltGDmin    & 15.8243 & 0.3246 & 0.5056 & 186.2968 \\
& L+S         & 15.9761 & 0.3306 & 0.4615 & 287.8737 \\
& Unet-frame & 15.9914 & 0.4898 & 0.3308 & 122.7593 \\
& Unet-seq   & 14.8047 & 0.3095 & 0.1755 & 52.2228 \\
& DVP         & 14.3000 & 0.4839 & 0.4072 & 156.0854 \\

& Proposed    & \textbf{17.0532} & \textbf{0.8796} &
\textbf{0.0910} & \textbf{40.7730} \\
\midrule

\multirow{6}{*}{\shortstack[c]{Non-rigid scene 1\\
(Fig.~\ref{fig:simulated_dynamic_results})}}
& AltGDmin    & 15.1408 & 0.3460 & 0.5407 & 187.9218 \\
& L+S         & 15.3174 & 0.3239 & 0.4389 & 282.6118 \\
& Unet-frame & 16.1944 & 0.4569 & 0.2940 & 105.6542 \\
& Unet-seq   & 16.1289 & 0.4603 & 0.2395 & 75.5565 \\
& DVP         & \textbf{18.3973} & 0.3669 & 0.1478 & 125.0182 \\

& Proposed    & 15.1407 & \textbf{0.9025} &
\textbf{0.1243} & \textbf{68.8088} \\
\midrule

\multirow{6}{*}{\shortstack[c]{Non-rigid scene 2\\
(Fig.~\ref{fig:simulated_dynamic_results2})}}
& AltGDmin    & 15.8571 & 0.3538 & 0.5473 & 163.8662 \\
& L+S         & 16.0726 & 0.3320 & 0.4123 & 229.5990 \\
& Unet-frame & \textbf{17.2473} & 0.3976 & 0.1843 & \textbf{80.4443} \\
& Unet-seq   & 17.1454 & 0.3445 & 0.2454 & 103.4324 \\
& DVP         & 16.6677 & 0.3534 & 0.3255 & 152.1201 \\

& Proposed    & 15.6098 & \textbf{0.8972} &
\textbf{0.1516} & 93.6818 \\
\bottomrule
\end{tabular}%
}
\end{table}

\subsection{Simulated Results}

First, we evaluate the proposed method using simulated data generated from the forward model in Eq.~\eqref{eqn:Forward}. The simulated data is generated using a wide FoV scene with a resolution of \(224 \times 112\) pixels. To approximate the noise characteristics of the mmWave imaging system, we add Gaussian noise with zero mean and a standard deviation of 0.078 to the scene, which is normalized to the range  [0,1]. The PSF is modeled as given by Eq.~\eqref{eq:psf_real}. \del{The narrow FoV measurements with a resolution of \(112 \times 112\) are generated by convolving the scene with the PSF and applying the masking function.} \add{
The narrow FoV measurements with a resolution of $112 \times 112$ are generated by convolving the wide-FoV scene with the PSF and \add{then applying the spatial cropping operator $\mathbf{M}$ to extract the central $112 \times 112$ region.}
}

We generate three noisy, narrow FoV sequences from wide FoV ground truths, each with 200 frames. In the first sequence, we translate a rigid human silhouette across the scene. In the second and third sequences, we depict non-rigid motion across the scene, with the silhouette's arms and legs moving as if running and walking. We apply our INR to each sequence to denoise the measurements and extend their FoV.  We compute the four evaluation metrics between the reconstructed wide FoV scene and the GT wide FoV scene. The quantitative results are summarized in Table~\ref{tab:sim_quant_results}. \del{Across all simulated scenarios, the proposed method consistently achieves the best reconstruction quality, outperforming all baselines in both LPIPS and FVD, demonstrating its ability to reconstruct the wide FoV scene with higher perceptual fidelity and better temporal coherence.}
\add{Across all simulated scenarios, the proposed method consistently achieves the highest SSIM and lowest LPIPS, indicating superior structural and perceptual fidelity. It also achieves the lowest FVD in two of the three sequences, demonstrating strong overall video reconstruction quality despite not consistently attaining the highest PSNR.}

Representative reconstructed frames from one of the non-rigid simulated scenes are presented in Fig.~\ref{fig:simulated_dynamic_results}. Corresponding results for the remaining two simulated scenes are provided in Appendix~\ref{sec:appendix b}, as shown in Fig.~\ref{fig:simulated_results} and Fig.~\ref{fig:simulated_dynamic_results2}.
These visual results demonstrate that the proposed method successfully reconstructs a wide FoV from narrow FoV measurements while effectively suppressing the aliasing artifacts.

\subsection{Experimental Results}
\label{subsec:experiments}

The dataset used in our experiments consists of passive-mode mmWave recordings of scenes with individuals moving across the FoV, all set against a static background. For our experiments, we focus on a scenario involving a single person moving across the FoV with 200 frames.

\begin{figure*}
    \centering
        \includegraphics[width=\textwidth]{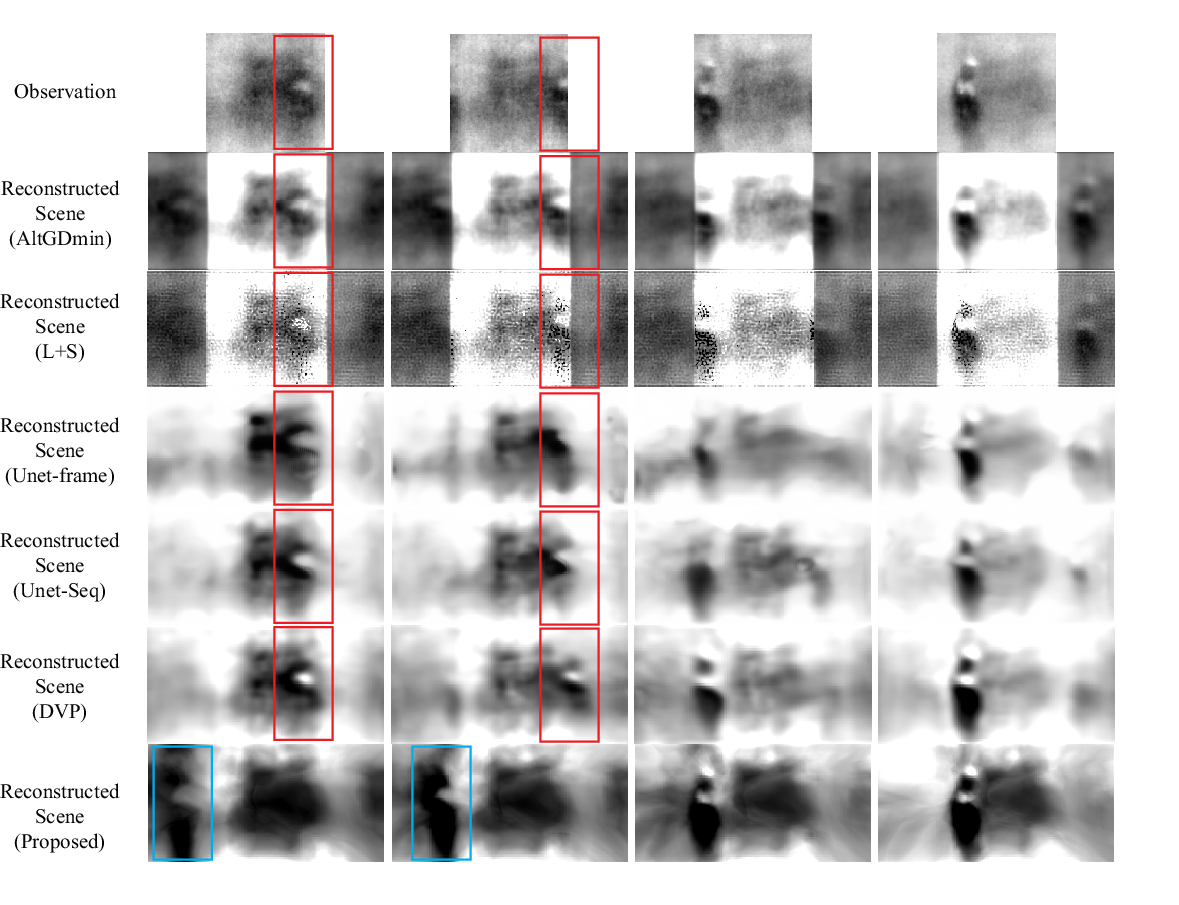}
        \caption{{\bf Extended FoV reconstruction of a real-world scene}. Selected reconstructed frames from the real-world mmWave measurements. The red bounding boxes highlight the aliasing artifacts in the narrow-FoV observations and the corresponding regions in the baseline reconstructions. While the baseline methods suppress these artifacts to varying degrees, the proposed method leverages the aliased information to recover the off-FoV person, highlighted by the blue bounding boxes. See the supplementary video for the complete sequences.}
        \label{fig:real_world_results}
\end{figure*}

As in the simulated experiments, the PSF for the real-world data is modeled using the formulation in Eq.~\eqref{eq:psf_real}. Selected reconstructed frames for the scenario involving a single moving person are shown in Fig.~\ref{fig:real_world_results}. These visual results demonstrate that the proposed method successfully reconstructs a wide FoV from narrow FoV measurements while effectively suppressing the aliasing artifacts. However, the reconstruction quality is lower compared to the simulated results. \del{This degradation arises from modeling mismatches between the analytical PSF used during reconstruction and the true forward operator of the hardware system. In practice, the real system exhibits effects that are not fully captured by the PSF model.}
\add{This degradation may be attributed to mismatches between the analytical forward model and the physical imaging process. First, although the measured noise is approximately Gaussian and spatially white, deviations from the assumed independent and identically distributed Gaussian noise model may affect reconstruction accuracy. Second, the analytical PSF incorporates aperture tapering and Gaussian blur but may not fully capture hardware imperfections, including nonuniform sensor responses, optical aberrations, and spatially varying blur. Finally, temporal variations in the experimental imaging conditions may introduce changes in the measurement process that are not captured by the fixed forward operator. These unmodeled effects may contribute to the reduced reconstruction quality observed in real-world measurements.} 
\del{Similar to the behavior seen in Fig.~\ref{fig:simulated_dynamic_results}, DVP performs well in regions without aliasing artifacts, but it fails to remove the artifact components when aliasing is present, whereas the proposed method remains robust.}
\add{Consistent with the simulated results in Fig.~\ref{fig:simulated_dynamic_results}, the baseline methods reconstruct the observed scene and suppress aliasing artifacts to varying degrees. However, their ability to recover information beyond the native FoV remains limited. In contrast, the proposed method more effectively reconstructs the out-of-FoV person from the aliased measurements, as highlighted by the blue bounding boxes.}

\add{
\subsection{Computational Cost and Runtime}
\label{subsec:runtime}

We evaluate the computational requirements of the proposed framework and baseline methods using real-world mmWave measurements. All experiments are conducted on an NVIDIA Tesla V100 GPU (32 GB), using 200-frame sequences with a reconstructed resolution of $224 \times 112$ pixels. Table~\ref{tab:computational_cost} summarizes the optimization time, reconstruction time, number of iterations or epochs, NN model parameters, and peak GPU memory consumption. The proposed INR requires only 100k parameters, substantially fewer than DVP (8.6M) and the U-Net baselines (13.4M). Furthermore, the proposed method reconstructs the complete 200-frame sequence in approximately 0.33 seconds after optimization, demonstrating efficient inference despite the computational cost of sequence-specific optimization.
}
\begin{table}[t]
\centering
\addtable
\caption{Computational cost comparison on real-world
mmWave data using an NVIDIA Tesla V100 GPU (32 GB).
All methods reconstruct 200 frames of resolution
$224 \times 112$. GPU memory denotes peak allocated
memory during optimization.}
\label{tab:computational_cost}
\setlength{\tabcolsep}{5.5pt}
\renewcommand{\arraystretch}{1.1}
\small
\begin{tabular}{@{}lrrrrr@{}}
\toprule
Method & Iter. & Params & Opt.
       & Rec. (s) & GPU (MiB) \\
\midrule
AltGDmin    & 50   & N/A   & 5.7 s
            & 0.0001 & 210.11 \\
L+S         & 6000 & N/A   & 4.48 h
            & 0.0001 & 237.84 \\
Unet-Frame & 6000 & 13.4M & 6.03 h
            & 1.1693 & 317.64 \\
Unet-Seq   & 6000 & 13.4M & 7.39 h
            & 1.0458 & 318.65 \\
DVP         & 76   & 8.6M  & 306 s
& 0.3094 & 192.06 \\
Proposed    & 6000 & 100k  & 4.52 h
            & 0.3294 & 230.74 \\
\bottomrule
\end{tabular}
\end{table}

\section{Conclusion}

In this work, we introduced a general framework for reconstructing wide FoV scenes from narrow FoV measurements \add{acquired by sparsely sampled real-aperture passive} mmWave imaging systems. Such systems, which often rely on discretized sampling during signal acquisition or optical upconversion, are susceptible to aliasing artifacts that complicate interpretation. Our approach leverages time-varying INRs to exploit temporal regularity across sequential frames, enabling accurate wide FoV reconstruction while substantially mitigating aliasing arising from discretized sampling. \add{Across all simulated scenarios, the proposed method consistently achieves the highest SSIM and lowest LPIPS, indicating superior structural and perceptual fidelity. It also achieves the lowest FVD in two of the three sequences, demonstrating strong overall video reconstruction quality, although it does not consistently attain the highest PSNR.}

We demonstrated the effectiveness of the proposed method using measurements from the mmWave imaging system developed by Phase Sensitive Innovations (PSI), showing that the framework can reconstruct a wide FoV from narrow FoV observations while suppressing aliasing artifacts. In addition, we performed an extensive characterization of the system noise, analyzing its statistical distribution, spatial and temporal correlations, and power spectral density to better understand the measurement characteristics that influence reconstruction quality.

While the method achieves promising results for moderate-length sequences, scaling to longer sequences or highly dynamic scenes remains an open area for future work. In addition, the current forward model does not fully capture the true system operator, which can lead to performance degradation when applying the framework to real-world data. Future work will focus on developing more accurate forward models that incorporate additional system effects to further improve reconstruction fidelity. \add{Furthermore, real-world validation is currently limited to a single experimental sequence due to data availability constraints. Although the simulation results demonstrate the effectiveness of the proposed framework across different motion patterns, additional real-world experiments involving diverse scenes and imaging conditions are needed to establish its generalizability.} Further, like many learning-based imaging pipelines, the current framework incurs notable computational overhead, rendering real-time application challenging. Advances in neural representation architectures, efficient optimization strategies, and hardware acceleration will be essential to enable near–real-time performance and broaden practical deployment across diverse mmWave imaging platforms. Taken together, our results highlight the potential of the proposed INR-based framework to enhance the capabilities of spatially aliased mmWave imaging systems, thereby broadening their applicability across a wide range of scenarios.

\section*{Acknowledgments}
We thank Phase Sensitive Innovations, Inc. for providing the mmWave imaging dataset used in this work. S.H.~was supported by the SMART Scholarship from the U.S. Department of Defense. C.A.M.~was supported in part by ONR contract/grant nos.~N6833522G0039 and~N00014-23-1-2752,  AFOSR Young Investigator Program award no.~FA9550-22-1-0208, and NSF CAREER award no.~2339616. This work was supported in part by high-performance computer time and resources from the DoD High Performance Computing Modernization Program.

The U.S. Government is authorized to reproduce and distribute reprints for governmental purposes notwithstanding any copyright notation thereon. The views and conclusions contained herein are those of the authors and should not be interpreted as necessarily representing the official policies or endorsements, either expressed or implied, of the Dept. of the Army or the Office of the Under Secretary of Defense for Research and Engineering (OUSD(R\&E)) or the U.S. Government. 

We thank Dr.~Joseph Mait for useful discussions about this work.

\appendices

\begin{figure*}[h!]
    \centering
        \includegraphics[width=\textwidth]{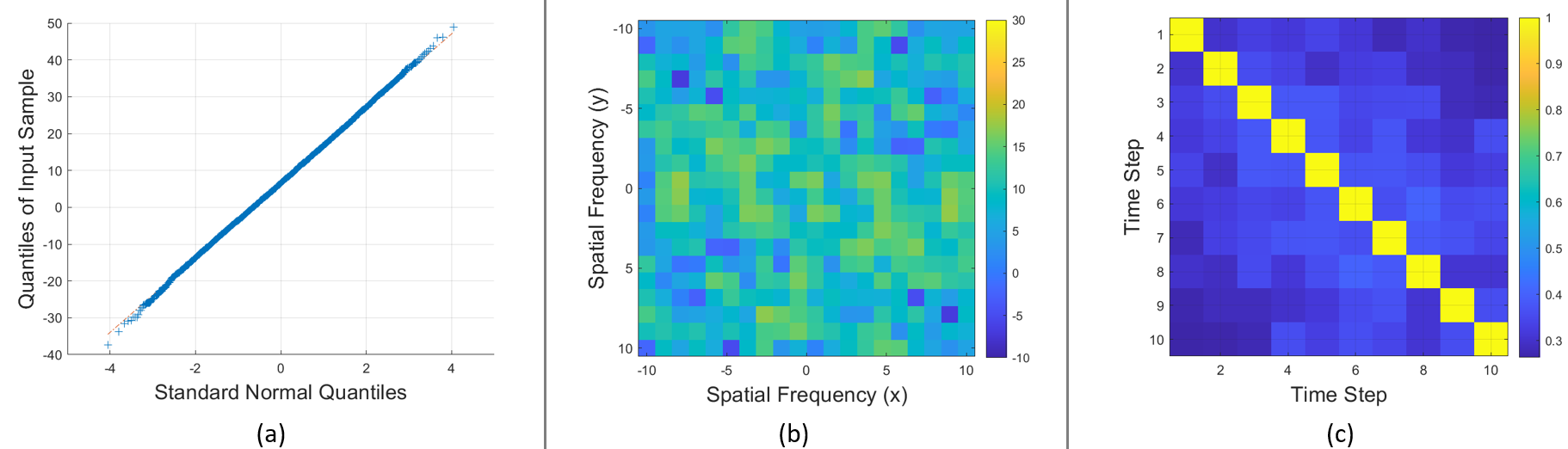}
        \caption{{\bf Characterization of the noise of the mmWave imaging system}. (a) QQ plot of the noise (residual) intensity values, indicating a Gaussian distribution. (b)  PSD (dB) of the noise intensity values, showing a broadly flat frequency concentration.  (c) Temporal covariance matrix of the noise intensity values, showing low correlation between adjacent time steps.  }
        \label{fig:noise_char}
\end{figure*}

\renewcommand{\theequation}{\thesection.\arabic{equation}} 
\setcounter{equation}{0} 

\section{Noise Characterization}
\label{sec:appendix a}

We characterize the mmWave system's noise using a measurement of a static scene to ensure that any temporal variation arises from the instrument rather than the scene. To avoid distorting the statistics, we work directly with the RAW frames obtained from the PSI system. This ensures that all display-oriented post-processing steps, such as filtering, per-frame min-max scaling, and bit-depth conversion, are bypassed, which can otherwise result in alteration to noise characteristics.  

We consider 200 RAW frames acquired under identical settings for the noise analysis. We estimate a background image by the temporal mean of the stack and subtract it from each frame to obtain residuals. For a static scene, this removes the deterministic scene content and fixed offsets, so that we can approximate the residual as the approximation of the system noise. Using these residuals as the system's noise approximation, we then assess the statistical distribution, power spectral density (PSD), and the temporal correlation of the noise.

\subsection{Statistical Distribution of Noise}

One fundamental aspect of noise characterization is to determine its statistical distribution. To assess whether the noise follows a Gaussian distribution, we generate a quantile-quantile (QQ) plot of the residual values of all the frames. QQ plots compare the quantiles of the computed residuals against the quantiles of a theoretical normal distribution. If the data follows a normal distribution, the points in the QQ plot will fall along a straight diagonal line. Any deviations from this line indicate departures from normality.

The computed QQ plot is shown in Fig.~\ref{fig:noise_char} (a). The plot demonstrates that the data points closely align with the diagonal, indicating that the noise in the system is approximately Gaussian distributed. This finding enables us to treat the noise as approximately Gaussian for the noise characterization.

\subsection{Noise Spatial Correlation}

We use PSD analysis to better understand the frequency content of the noise. The PSD is estimated using an extension of the Welch method \cite{Welch1967} adapted for image data. This method involves dividing the image into overlapping blocks, applying a window function to each block, computing the Fourier transform of each block, and averaging the squared magnitudes of the Fourier transform to obtain the final PSD estimate. For computing the Welch method for each residual frame, we considered a $20\times20$ pixel block and used a Hamming window with an overlap of 50\% between adjacent blocks.

Fig.~\ref{fig:noise_char}(b) shows the PSD of a representative residual frame. Aside from small, frequency-localized peaks and dips, the spectrum is broadly flat across the spectrum. This same behavior holds across frames, indicating that spectral coloring is weak and the noise can be modeled as approximately white. The minor deviations from flatness are consistent with windowing and imperfect background removal rather than true structured noise.

\subsection{Noise Temporal Correlation}

In addition to the above assessments, we also assess the temporal dependence in the noise. By temporal dependence, we mean the dependence between noise samples separated in time at the same spatial region. To quantify it, we select 36 non-overlapping $1\times10$ pixel patches across the image. For each patch, we vectorize the patch for 10 successive frames, and stack these 10-element vectors as columns of a matrix and compute the correlation matrix across those columns. Fig.~\ref{fig:noise_char}(c) shows a representative temporal correlation matrix for one patch. Off-diagonal entries are small—even at adjacent time steps, indicating weak inter-frame dependence. Taken together across patches, these results suggest the system noise is low in temporal correlation or approximately temporally white. Minor residual correlations can arise from slight background-estimation errors.

\section{Additional simulation results}
\label{sec:appendix b}

\begin{figure}
    \centering
        \includegraphics[width=.5\textwidth]{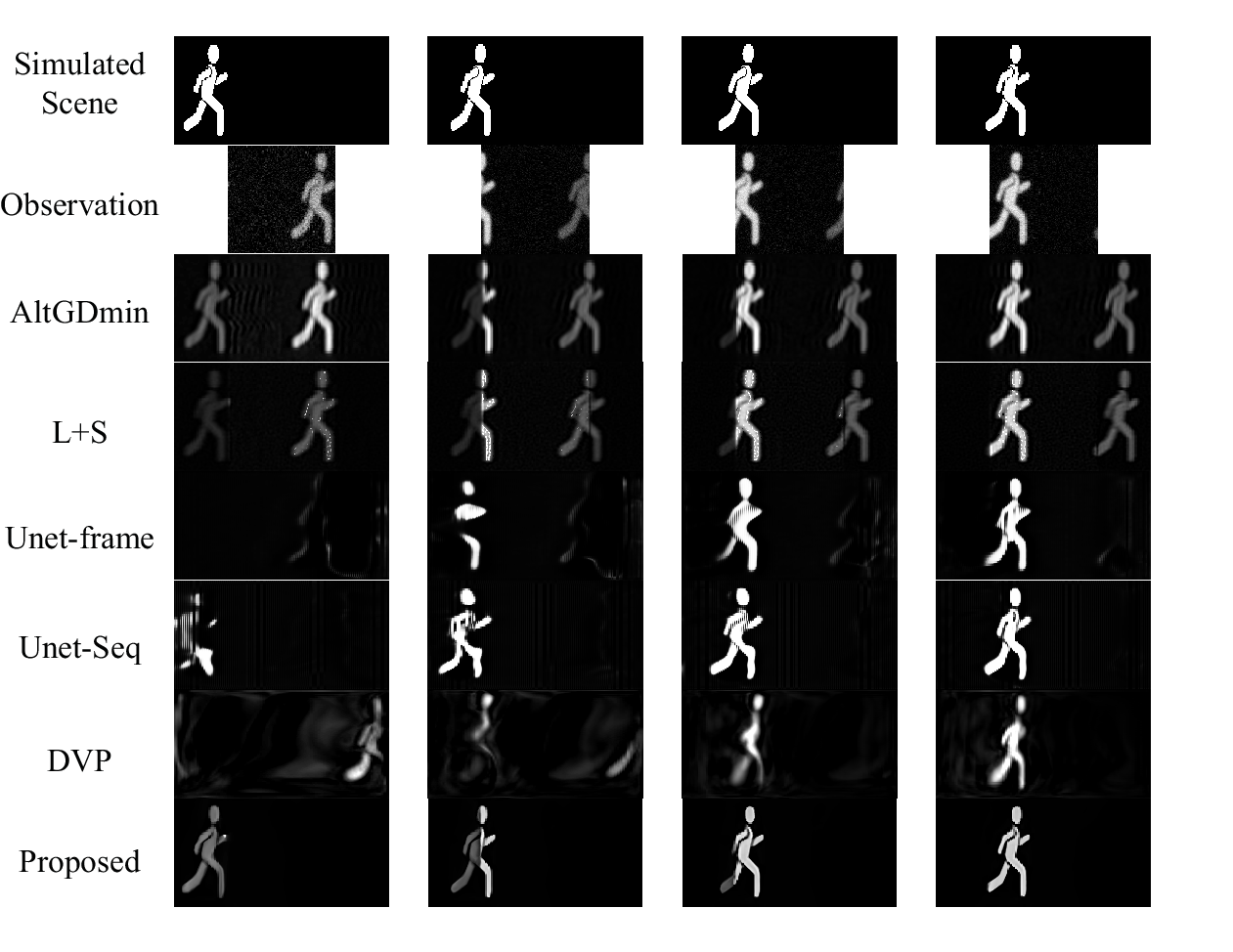}
        \caption{{\bf Extended FoV reconstruction of simulated rigid scene}. For the rigid scene, the proposed method recovers the full field of view from narrow FoV measurements and suppresses aliasing artifacts. Although some baseline methods partially suppress aliasing, their reconstructions exhibit residual artifacts or structural distortions, resulting in reduced reconstruction fidelity.}
        \label{fig:simulated_results}
\end{figure}

\begin{figure}
    \centering
        \includegraphics[width=.5\textwidth]{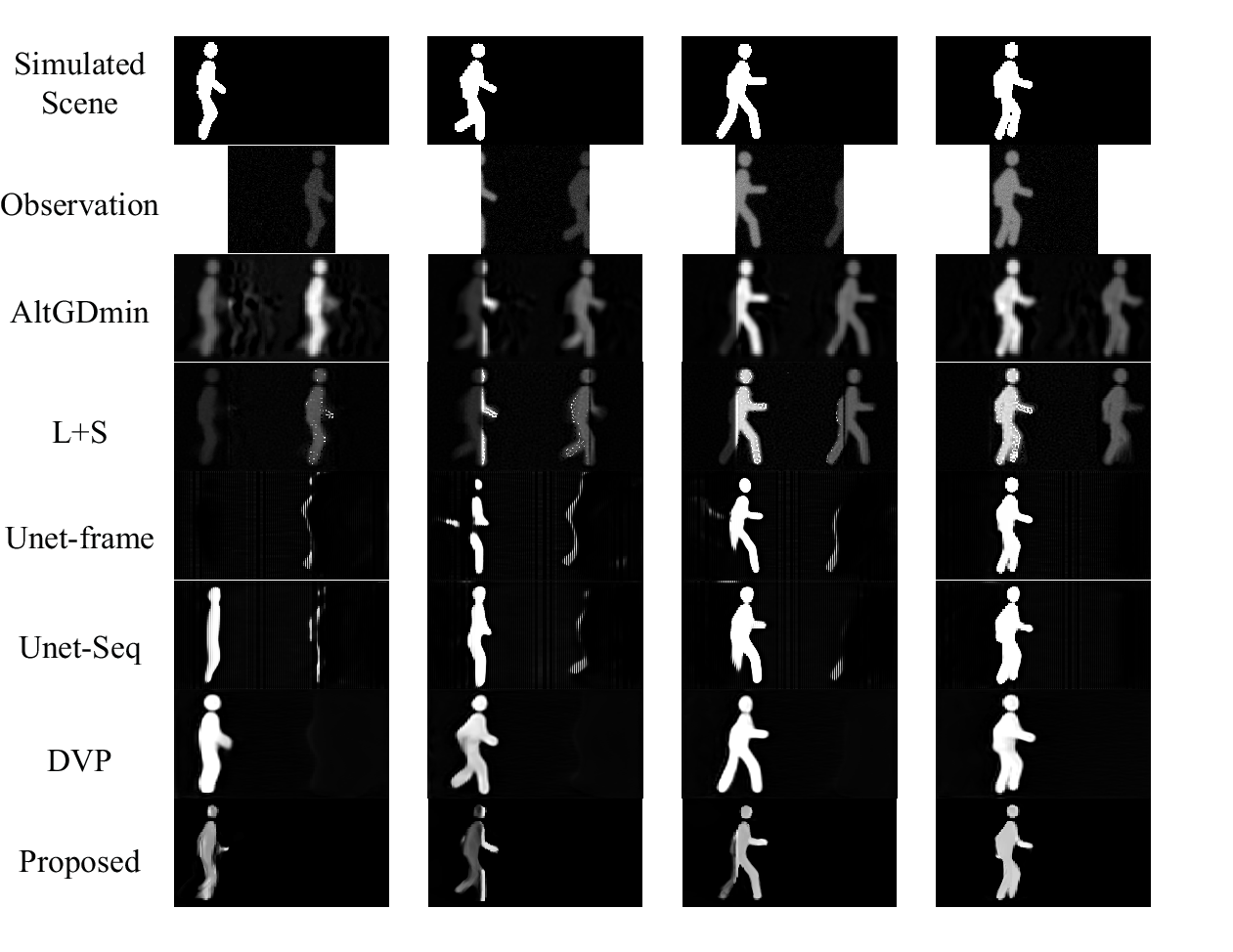}
        \caption{{\bf Extended FoV reconstruction of simulated non-rigid scene (walking motion)}. A second non-rigid scenario demonstrating robustness across different motion patterns. The proposed method yields cleaner reconstructions and successful artifact suppression across frames. DVP also achieves good reconstruction quality with minimal residual artifacts, whereas the remaining baselines exhibit incomplete scene recovery, residual aliasing, or structural distortions.}
        \label{fig:simulated_dynamic_results2}
\end{figure}

In addition to the non-rigid scenario presented in Fig.~\ref{fig:simulated_dynamic_results}, reconstructed frames from additional simulated scenes are provided in Fig.~\ref{fig:simulated_results} and Fig.~\ref{fig:simulated_dynamic_results2}. 
Together, these results demonstrate the robustness of the proposed framework across different simulated conditions.

\section{Implementation Details of Baseline Methods}
\label{sec:supp_baselines}

\subsection{Deep Video Prior (DVP)}

We employ the PyTorch implementation of Deep Video Prior (DVP) provided in~\cite{dvp_git}. DVP extends the Deep Image Prior (DIP) framework~\cite{ulyanov2018deep} to video reconstruction by exploiting temporal coherence across frames. Rather than reconstructing each frame independently, DVP optimizes a CNN over the entire video sequence.

We modify the implementation to use the same PSF and masking operator as in our forward model so that the network outputs wide FoV reconstructions consistent
with the observed narrow FoV measurements. The network is optimized independently for each observed sequence.

\add{
\subsection{Low-Rank Reconstruction via AltGDmin}

The low-rank baseline assumes that the underlying wide-FoV video scene can be approximated by a low-rank matrix. Given the linear forward operator $F(\cdot)$ and
measurement matrix $\mathbf{Y}$, the reconstruction problem is formulated as
\begin{equation}
    \min_{\mathbf{U},\mathbf{B}}
    \sum_{t=1}^{T}
    \left\|
    F(\mathbf{U}\mathbf{b}_t)-\mathbf{y}_t
    \right\|_2^2,
\end{equation}
where $\mathbf{X}=\mathbf{U}\mathbf{B}$,
$\mathbf{U}\in\mathbb{R}^{(H_xW_x)\times r}$,
$\mathbf{B}\in\mathbb{R}^{r\times T}$, and
$r\ll\min(H_xW_x,T)$.

We employ an AltGDmin-based alternating optimization strategy, adapted to our imaging forward operator $F(\mathbf{x})=\mathbf{M}(h*\mathbf{x})$.
With $\mathbf{U}$ fixed, $\mathbf{B}$ is obtained by solving the least-squares problem
\begin{equation}
    \mathbf{B}
    =
    \arg\min_{\mathbf{B}}
    \left\|
    \mathbf{\Phi}\mathbf{B}-\mathbf{Y}
    \right\|_F^2,
\end{equation}
where
$\mathbf{\Phi}=
[F(\mathbf{u}_1),\ldots,F(\mathbf{u}_r)]$
contains the forward projections of the columns of $\mathbf{U}$, expressed as vectors. With $\mathbf{B}$ fixed, $\mathbf{U}$ is updated using Adam, followed by QR orthonormalization.

\subsection{Low-Rank Plus Sparse Model (L+S)}

 The L+S baseline models the wide-FoV sequence as the sum of a low-rank component \(\mathbf{L}\), representing temporally correlated scene content, and a sparse component \(\mathbf{S}\), capturing localized variations due to motion. Adapting the L+S formulation to our PSF-based imaging operator, we solve
\begin{equation}
    \min_{L,S}\;
    \frac{1}{2}\left\|F(L+S)-Y\right\|_F^2
    +\lambda_L\|L\|_*
    +\lambda_S\|S\|_1.
    \label{eq:ls_baseline}
\end{equation}

where \(F\) denotes the same PSF-based forward operator used in the proposed framework, and \(\|\cdot\|_*\) denotes the nuclear norm. We solve this optimization problem using proximal gradient descent with singular-value thresholding and elementwise soft-thresholding. The regularization parameters \(\lambda_L=1\) and \(\lambda_S=0.01\) were selected through a grid search.
}
\subsection{U-Net Baselines}

We implement two U-Net-based reconstruction models to compare frame-wise
and temporally informed CNN reconstruction.

\paragraph{Unet-Frame}
The Unet-Frame model reconstructs each wide-FoV frame from its corresponding
narrow-FoV measurement:
\begin{equation}
    \hat{\mathbf{x}}_t
    =g_{\phi}(\mathbf{y}_t),
\end{equation}
where $g_{\phi}$ denotes the U-Net with trainable parameters $\phi$.
The network receives a single frame as input without explicitly incorporating
neighboring frames.

\paragraph{Unet-Seq}
The Unet-Seq model incorporates temporal context by using a sequence of $L$
consecutive narrow-FoV frames to reconstruct the final frame in the input window:
\begin{equation}
    \hat{\mathbf{x}}_t
    =
    g_{\psi}(
    \mathbf{y}_{t-L+1},
    \ldots,
    \mathbf{y}_t),
\end{equation}
where $g_{\psi}$ denotes the sequence-based U-Net. We use $L=10$ in our experiments.

\add{
Both U-Net variants are trained from scratch independently for each test sequence.
}

\section{Ablation Studies}

\add{
\subsection{PSF Parameter Selection}
\label{subsec:psf_ablation}

To evaluate the sensitivity of the proposed framework to the approximate PSF, we conduct an ablation study using real-world mmWave measurements. We consider three Kaiser window parameters, $\beta \in \{2,4,6\}$, and three Gaussian blur standard deviations, $\sigma_{\mathrm{G}} \in \{0.5,1.0,1.5\}$, resulting in nine PSF configurations. Figure~\ref{fig:psf_parameter_comparison} presents the corresponding reconstructions.

\begin{figure}[t]
    \centering
    \addtable

    \includegraphics[width=0.32\columnwidth]{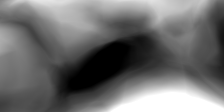}\hfill
    \includegraphics[width=0.32\columnwidth]{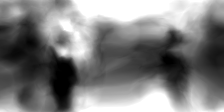}\hfill
    \includegraphics[width=0.32\columnwidth]{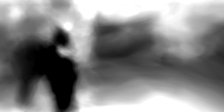}

    \vspace{2pt}

    \includegraphics[width=0.32\columnwidth]{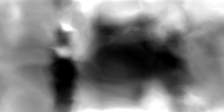}\hfill
    \includegraphics[width=0.32\columnwidth]{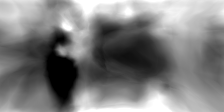}\hfill
    \includegraphics[width=0.32\columnwidth]{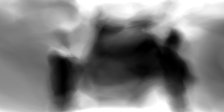}

    \vspace{2pt}

    \includegraphics[width=0.32\columnwidth]{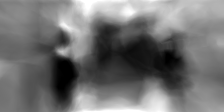}\hfill
    \includegraphics[width=0.32\columnwidth]{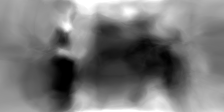}\hfill
    \includegraphics[width=0.32\columnwidth]{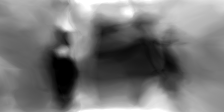}

    \caption{\textbf{PSF parameter sensitivity.}
    Reconstruction results for different Kaiser
    window parameters $\beta$ (rows: 2, 4, 6)
    and Gaussian blur standard deviations
    $\sigma_G$ (columns: 0.5, 1, 1.5).
    The combination $\beta=4$, $\sigma_G=1$
    provides the best qualitative result
    among the tested settings.}

    \label{fig:psf_parameter_comparison}
\end{figure}

Among the tested configurations, $\beta=4$ and $\sigma_{\mathrm{G}}=1$ provide the most favorable qualitative reconstruction, effectively suppressing visible aliasing artifacts while preserving scene details. These results motivate the parameter values used in our experiments. Nevertheless, the selected parameters represent an empirical approximation and do not constitute a direct physical calibration of the imaging system.
}

\begin{figure}[t]
    \centering
    \addtable
    \includegraphics[width=\linewidth]{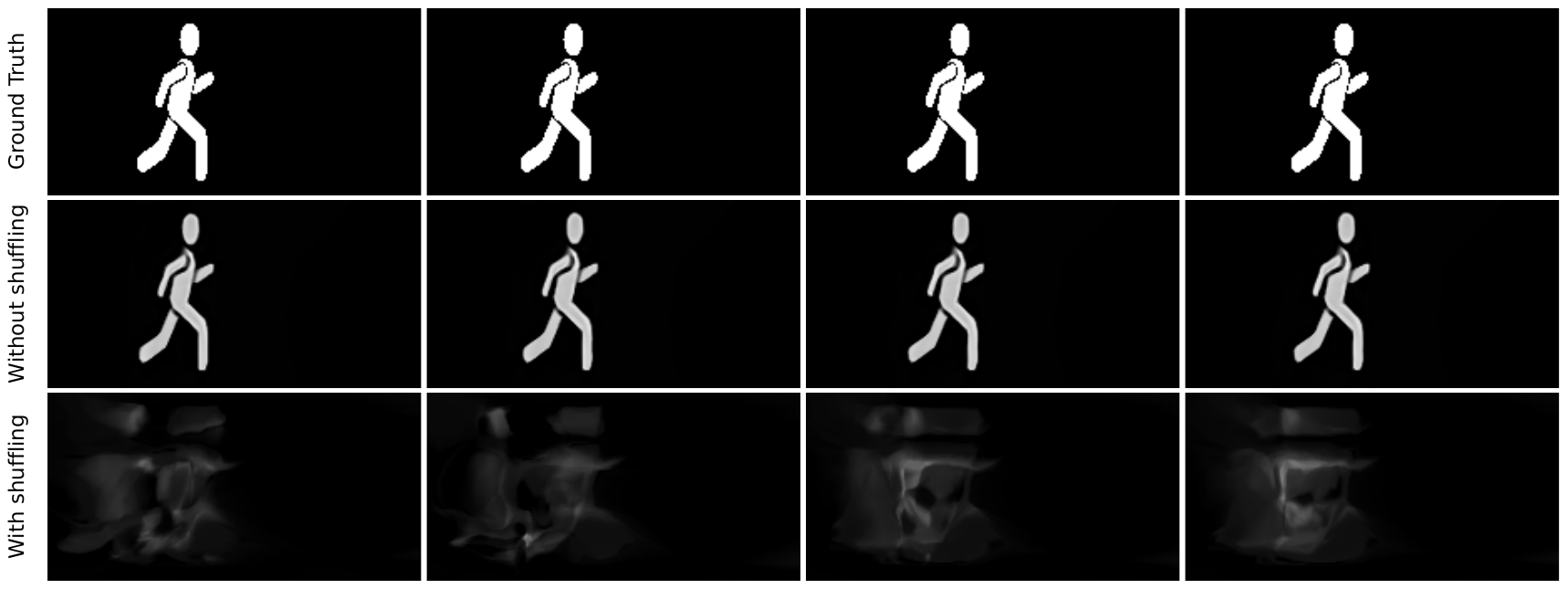}

    \caption{\textbf{Effect of temporal shuffling on INR reconstruction.}
    Four representative frames showing ground truth (top), reconstructions without shuffling (middle), and with shuffling (bottom).
    Temporal shuffling significantly degrades reconstruction quality, highlighting the importance of temporal coherence in the proposed INR framework.}
    
    \label{fig:temporal_shuffling}
\end{figure}

\add{
\subsection{Effect of Temporal Ordering}
\label{subsec:temporal_ablation}

To evaluate whether the temporal input contributes meaningfully to the reconstruction, we conduct an ablation study that disrupts the temporal structure of the training sequence. Specifically, we randomly shuffle the temporal order of the input frames while keeping the frame content, network architecture, and training procedure otherwise unchanged. This disrupts the natural temporal correlations between consecutive measurements.

Figure~\ref{fig:temporal_shuffling} compares reconstructions obtained using the original temporal ordering and the shuffled ordering. The reconstructions obtained from the shuffled sequence exhibit visibly degraded quality, with reduced consistency and weaker recovery of scene structure. This result indicates that the proposed INR does not treat the time index as a purely independent auxiliary coordinate. Instead, the shared spatio-temporal parameterization benefits from the temporal structure present in the original sequence.

}

\begin{figure}[t]
    \centering
    \addtable
    \includegraphics[width=\linewidth]{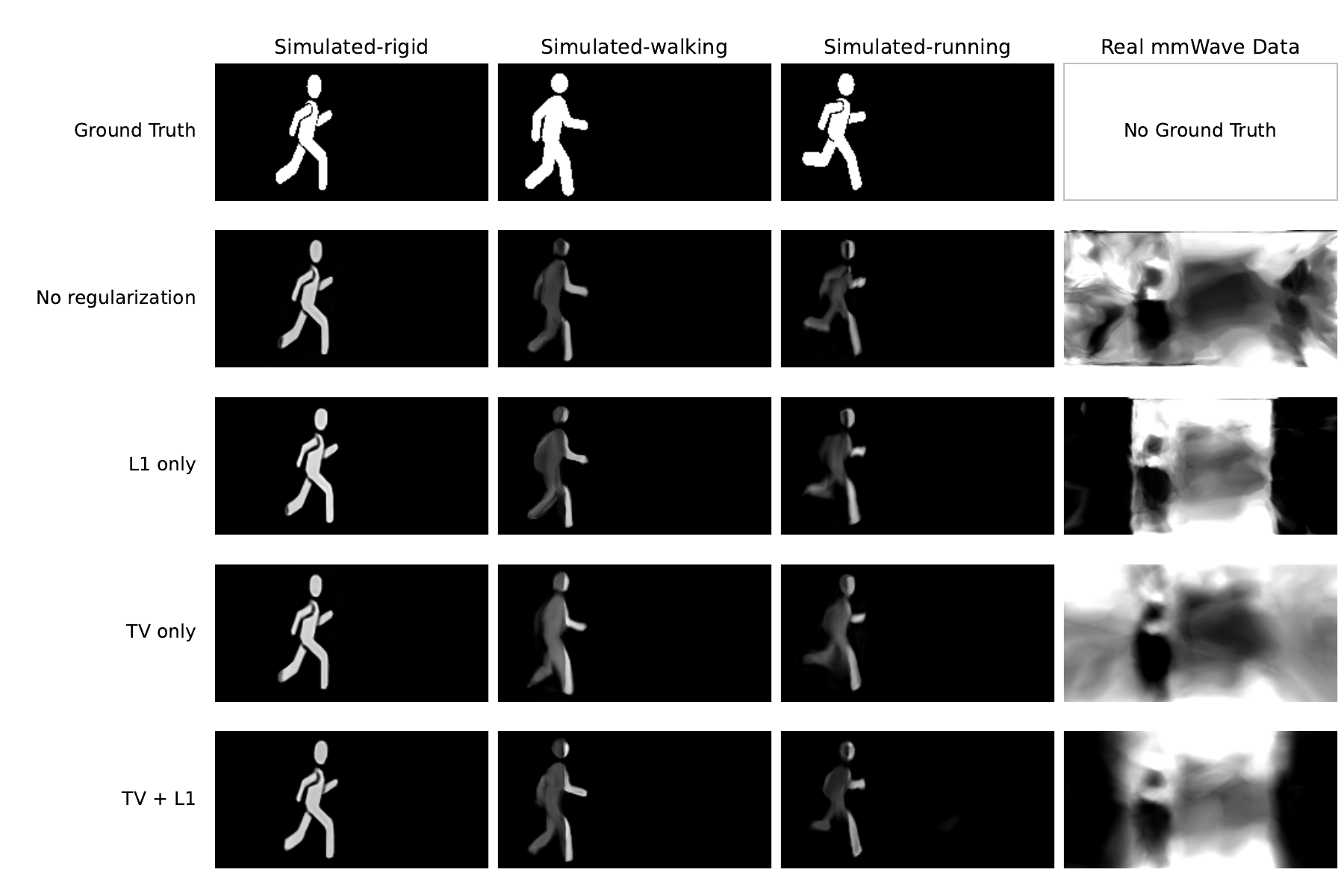}
    \caption{\textbf{Qualitative comparison of regularization
    strategies.} The first row shows the ground-truth simulated scenes,
    while the remaining rows show INR reconstructions
    without regularization, with $L_1$ regularization,
    with TV regularization, and with both $L_1$ and TV
    regularization. }
    \label{fig:regularization_ablation}
\end{figure}

\add{
\subsection{Effect of Explicit Regularization}
\label{subsec:reg_ablation}

To investigate the effect of explicit regularization on INR-based reconstruction, we conduct an ablation study using four configurations: no regularization, L1 regularization only, TV regularization only, and both L1 and TV regularization. Table~\ref{tab:reg_ablation} presents the quantitative results averaged over three simulated sequences. The results indicate that explicit regularization does not consistently improve reconstruction quality, suggesting that the implicit prior imposed by the INR is sufficient for the simulated scenarios.

However, qualitative evaluation of experimental mmWave measurements reveals that TV regularization helps suppress residual artifacts in the reconstructed scenes, as illustrated in Fig.~\ref{fig:regularization_ablation}. This may be attributed to its ability to promote spatial smoothness while preserving object boundaries, thereby reducing artifacts associated with measurement noise and forward-model inaccuracies. These findings suggest that while the implicit regularization provided by the INR is sufficient for simulated data, additional spatial regularization can be beneficial for experimental measurements. Therefore, we augment the original log-likelihood loss with a TV regularization term.

\begin{table}[t]
\centering
\addtable
\caption{Regularization ablation study averaged over
three simulated sequences.}
\label{tab:reg_ablation}
\begin{tabular}{lcccc}
\toprule
Regularization & PSNR $\uparrow$ &
SSIM $\uparrow$ &
LPIPS $\downarrow$ &
FVD $\downarrow$ \\
\midrule
None &
15.859 & 0.8921 & \textbf{0.1242} & 70.445 \\
$L_1$ only &
15.344 & \textbf{0.8985} & 0.1305 & \textbf{67.044} \\
TV only &
\textbf{16.299} & 0.8936 & 0.1559 & 75.612 \\
$L_1$ + TV &
15.261 & 0.8942 & 0.1419 & 70.072 \\
\bottomrule
\end{tabular}
\end{table}

}

\add{
\subsection{Comparison of INR Architectures}
\label{subsec:inr_ablation}

To investigate the influence of INR architecture on reconstruction performance, we compare the proposed ReLU-based MLP with three alternatives designed to better represent high-frequency features: a Fourier-feature MLP~\cite{tancik2020fourier}, SIREN~\cite{sitzmann2020implicit}, and WIRE~\cite{saragadam2023wire}. All architectures use eight fully connected layers with 128 hidden units per layer and a skip connection after the fourth layer. Architecture-specific hyperparameters are optimized through a grid search. We evaluate the models on three simulated sequences and real-world mmWave measurements.

Fig.~\ref{fig:activation_comparison} and Table~\ref{tab:activation_comparison} present the qualitative and quantitative comparisons, respectively. Although SIREN and WIRE achieve higher PSNR and SSIM, respectively, the ReLU-based INR achieves the lowest LPIPS and FVD. The qualitative results further show that the ReLU-based INR produces coherent reconstructions across both simulated and real-world data, while the alternative architectures exhibit varying degrees of reconstruction artifacts. These findings suggest that increased high-frequency representation capacity does not necessarily improve reconstruction quality and may increase susceptibility to overfitting noise and artifacts in our underdetermined inverse problem. In contrast, the spectral bias of the ReLU-based MLP may provide beneficial implicit regularization by discouraging the fitting of unsupported high-frequency components.

\begin{figure}[t]
    \centering
    \addtable
    \includegraphics[width=\linewidth]{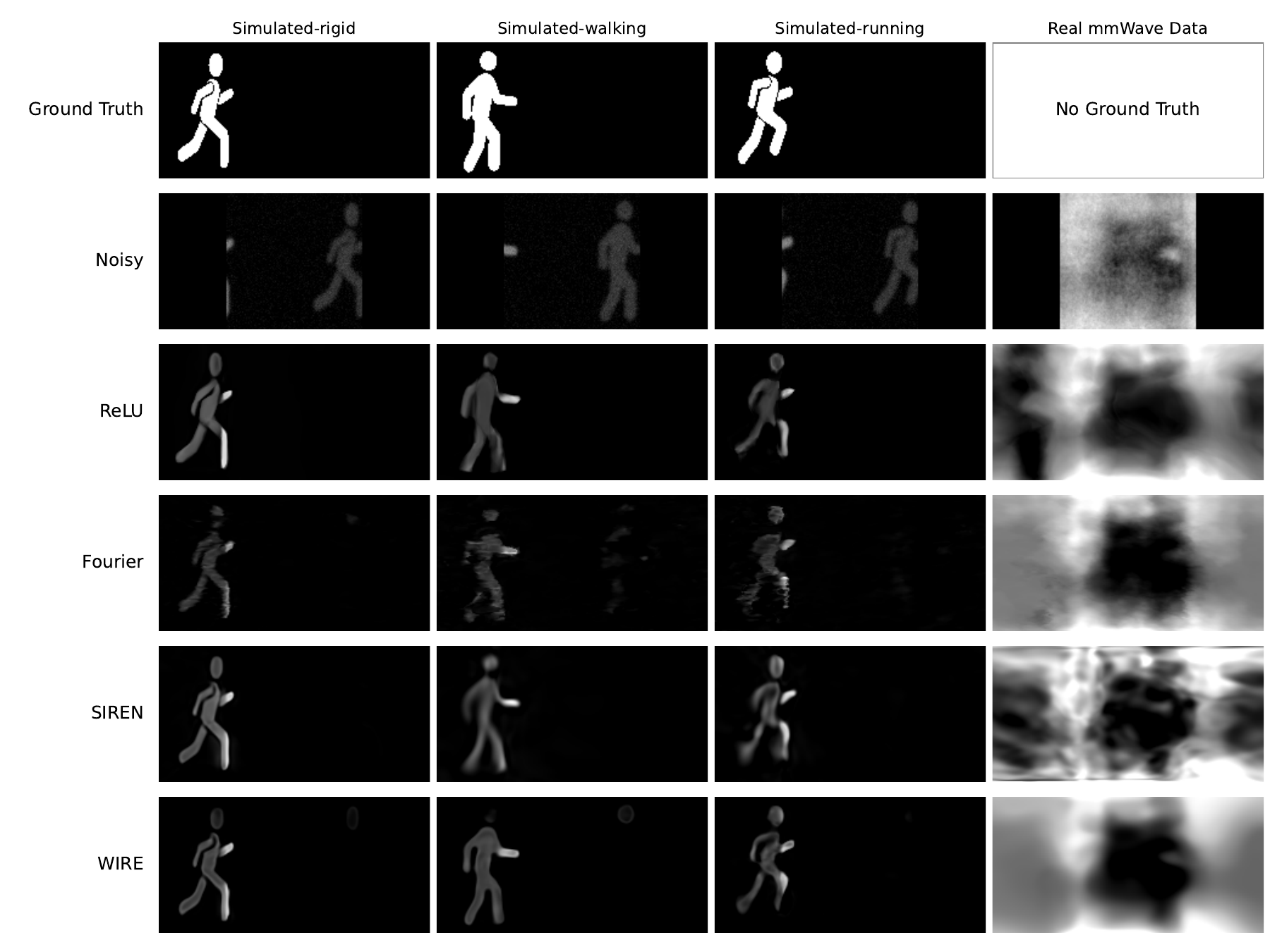}
    \caption{\textbf{Qualitative comparison of INR architectures for extended FoV reconstruction.} Representative frames using ReLU, Fourier features, SIREN, and WIRE on three simulated scenes and real-world passive mmWave data.}
    \label{fig:activation_comparison}
\end{figure}

\begin{table}[h]
    \centering
    \addtable
    \caption{Quantitative comparison of different INR architectures
    on simulated datasets. Best results are highlighted in bold.}
    \label{tab:activation_comparison}
    \begin{tabular}{lcccc}
        \toprule
        Method & PSNR $\uparrow$ & SSIM $\uparrow$
        & LPIPS $\downarrow$ & FVD $\downarrow$ \\
        \midrule
        ReLU    & 15.8592 & 0.8921 & \textbf{0.1242} & \textbf{70.4450} \\
        Fourier & 14.1723 & 0.8003 & 0.2885 & 173.0080 \\
        SIREN   & \textbf{16.4326} & 0.8514 & 0.1517 & 88.9744 \\
        WIRE    & 16.0661 & \textbf{0.9066} & 0.1348 & 82.6671 \\
        \bottomrule
    \end{tabular}
\end{table}

}

\bibliographystyle{IEEEtran}
\bibliography{refs}

\vspace{11pt}

\begin{IEEEbiographynophoto}{Sanjaya Herath}
received the B.S. degree in electrical and electronics engineering from the University of Peradeniya, Sri Lanka, in 2020. He is currently pursuing the Ph.D. degree in Electrical and Computer Engineering at the University of Maryland, College Park, MD, USA. His research interests include computational imaging, machine learning, and signal processing. He is a member of the UMD Intelligent Sensing Laboratory and received the SMART Scholarship from the U.S. Department of Defense in 2024. 
\end{IEEEbiographynophoto}

\begin{IEEEbiographynophoto}{Matthew R. Ziemann}
    received the B.S. degree in physics from the University of Illinois Urbana-Champaign, Urbana, IL, USA, in 2016, and the M.S. and Ph.D. degrees in computer science from the University of Maryland (UMD), College Park, MD, USA, in 2023 and 2026, respectively. His doctoral research was conducted with the UMD Intelligent Sensing Laboratory.
    He is currently a physicist with the DEVCOM Army Research Laboratory, Adelphi, MD, USA. His research focuses on the intersection of physics and deep learning, with applications in sensing.
\end{IEEEbiographynophoto}

\begin{IEEEbiographynophoto}{Christopher A. Metzler}
    received B.S., M.S., and Ph.D. degrees in electrical and computer engineering from Rice University, Houston, TX, USA, in 2013, 2014, and 2019, respectively.
    He was a Post-Doctoral Researcher at the Stanford Computational Imaging Laboratory, Stanford, CA, USA, from 2019 to 2020. He is an Associate Professor of computer science with the University of Maryland (UMD), College Park, MD, USA, where he directs the UMD Intelligent Sensing Laboratory. His research develops new systems and algorithms for solving problems in computational imaging and sensing, machine learning, and wireless communications.
    Dr. Metzler was an Intelligence Community Post-Doctoral Research Fellow, a NSF Graduate Research Fellow, a DoD NDSEG Fellow, and a NASA Texas Space Grant Consortium Fellow. His work has received multiple best paper awards. He recently received the NSF CAREER Award, the AFOSR Young Investigator Program Award, and the ARO Early Career Program Award.
\end{IEEEbiographynophoto}
\vfill

\end{document}